\documentclass[preprint,showpacs,amsmath,amssymb]{revtex4}
\usepackage{graphicx}
\usepackage{dcolumn}
\usepackage{bm}
\usepackage{textcomp}
\usepackage{epsfig}
\usepackage{color}
\newcommand{\beq}{\begin{equation}}
\newcommand{\eeq}{\end{equation}}
\newcommand{\bea}{\begin{eqnarray}}
\newcommand{\eea}{\end{eqnarray}}
\newcommand{\non}{\nonumber}

\newcommand{\bary}{\begin{array}}
\newcommand{\eary}{\end{array}}
\newcommand{\benum}{\begin{enumerate}}
\newcommand{\eenum}{\end{enumerate}}
\newcommand{\bitem}{\begin{itemize}}
\newcommand{\eitem}{\end{itemize}}
\begin{document}
\title{Strong-coupling expansion for the momentum distribution of the Bose Hubbard model with benchmarking against exact numerical results}

\author{J. K. Freericks$^{1}$,
H. R. Krishnamurthy$^{1,2,3}$,
Yasuyuki Kato$^4$,  Naoki Kawashima$^4$,
and Nandini Trivedi$^5$}
\affiliation{$^1$Department of Physics, Georgetown University,
 37th and O Sts. NW, Washington, DC 20057, USA\\
 $^2$Centre for Condensed Matter Theory, Department of Physics, Indian Institute of Science,
 Bangalore 560012, India\\
$^3$Condensed Matter Theory Unit,
Jawaharlal Nehru Centre for Advanced Scientific Research,
Bangalore 560064, India\\
$^4$Institute for Solid State Physics, University of Tokyo, Kashiwa, Chiba 277-8581, Japan\\
$^5$Department of Physics, The Ohio State University, Columbus, OH 43210, USA\\
}
\date{\today}

\begin{abstract}
A strong-coupling expansion for the Green's functions, self-energies and correlation functions of the Bose Hubbard model is developed. We illustrate the general formalism, which includes all possible inhomogeneous effects in the formalism, such as disorder, or a trap potential, as well as effects of thermal excitations. The expansion is then employed to calculate the momentum distribution of the bosons in the Mott phase for an infinite homogeneous periodic system at zero temperature through third-order in the hopping.  By using scaling theory for the critical behavior at zero momentum and at the critical value of the hopping for the Mott insulator to superfluid transition along with a generalization of the RPA-like form for the momentum distribution, we are able to extrapolate the series to infinite order and produce very accurate quantitative
results for the momentum distribution in a simple functional form for one, two, and three dimensions; the accuracy is better in higher dimensions and is on the order of a few percent relative error everywhere except close to the critical value of the hopping divided by the on-site repulsion. In addition, we find simple phenomenological expressions for the Mott phase lobes in two and three dimensions which are much more accurate than the truncated strong-coupling expansions and any other analytic approximation we are aware of. The strong-coupling expansions and scaling theory results are benchmarked against numerically exact QMC simulations in two and three dimensions and against DMRG calculations in one dimension. These analytic expressions will be useful for quick comparison of experimental results to theory and in many cases can bypass the need for expensive numerical simulations.
\end{abstract}
\maketitle

\section{Introduction}

The Bose Hubbard model~\cite{hubbard,fisher_etal,sheshadri_etal} was heavily studied as a simple model for disordered superconductors~\cite{fisher_etal}; subsequently it was demonstrated~\cite{jaksch} that ultra-cold atoms trapped in optical lattices provide an alternate, and more controllable, experimental realization of it, sparking even more interest.
One of the most useful tools for analyzing the states of cold atom systems is a time-of-flight measurement
of their momentum distribution when the lattice and trapping potentials are rapidly shut off and the atomic cloud is allowed to expand and then is imaged with absorption spectroscopy~\cite{greiner}.  The time-of-flight image, in the long-expansion-time limit, is directly related to the momentum distribution function of the atoms in the optical lattice before expansion~\cite{prokofiev_preprint,nandini_preprint}.

Even before cold atom systems were employed to examine Bose Hubbard model physics, the phase diagram of the model
was accurately determined in a strong-coupling approximation~\cite{strong1,strong2} (for a recent review of this early history, see Ref.~\onlinecite{weichman}).  This approach, which relied on expanding the properties in a perturbative series in the hopping, captured much of the behavior of the model, and when extrapolated via a scaling theory ansatz for the critical behavior at the tips of the Mott lobes~\cite{fisher_etal}, proved to be as accurate as the quantum Monte Carlo (QMC) simulations that had been performed at that time~\cite{scalettar,trivedi}.  Since then, the strong coupling perturbation theory has been pushed to higher order~\cite{elstner1,elstner2,one-d}, and the QMC simulations have improved dramatically in two~\cite{prokofiev_2d} and three dimensions~\cite{prokofiev_3d,nandini2}.  In addition, highly accurate density matrix renormalization group (DMRG) studies have been performed on the model in one dimension~\cite{krish_dmrg,kuhner_dmrg1,kuhner_dmrg2}.

Surprisingly, despite all of the work that has been performed on the phase diagrams with a strong-coupling
analysis, there are only limited results for the momentum distribution functions.  The first few terms of the structure factor have been determined to high order in one dimension~\cite{one-d} and the zero momentum distribution function has been examined in one and two dimensions~\cite{elstner2}. A recent random phase approximation (RPA) has been carried out~\cite{rpa}, which corresponds to the exact solution for the momentum distribution in the infinite-dimensional limit (see also Ref.~\onlinecite{nandini_rpa}).  In this contribution, we present an alternative formulation of the strong-coupling perturbation theory for the many-body Green's functions, which can be immediately employed to evaluate the momentum distribution function as a power series in the hopping divided by the interaction strength for each value of the momentum. Recently a similar strong-coupling formalism to ours has been proposed~\cite{pelster1} and used to calculate the momentum distribution in three dimensions through second order~\cite{pelster2}. We take our strong-coupling expansion and, guided by the exact solution from the RPA, {\it we construct an ansatz for the scaling behavior of the momentum distribution function and then employ it to produce analytic expressions for the momentum distribution that are accurate for all values of the hopping within the Mott phase}. These results could prove useful as a simple means to check against experimental data on more recent Bose Hubbard model systems~\cite{bloch,spielman1,spielman2}. We also take the results for the scaling behavior of the momentum distribution and use it as a phenomenological ansatz for the scaling behavior of the phase diagram that sums many more terms than the original ansatz.  Comparing that result with the QMC data in two and three dimensions also shows excellent agreement.

We write the bosonic Hubbard Hamiltonian in the presence of a potential in the form,
\bea
\nonumber
\mathcal{H} & = & \mathcal{H}_0 + \mathcal{H}_{hop} = \sum_j \mathcal{H}_{0 j} + \mathcal{H}_{hop}\\
\mathcal{H}_{0 j}& \equiv & [V_T(\mathbf{r}_j)\ -\mu]\hat{n}_j + \frac{U}{2} \hat{n}_j(\hat{n}_j-1)\\
\mathcal{H}_{hop}& \equiv & - \sum_{j,j^\prime} \mathfrak{t}_{j j^\prime} a^{\dagger}_j a_{j^\prime}
\label{eq: hubham}
\eea
Here $j, j^\prime$ label the sites of a (hypercubic) lattice in $d$ dimensions, with a lattice constant which we set equal to $1$ (the unit of distance); $\mathbf{r}_j$ is the position vector of the $j^{th}$ site as measured
from the center of the system. The symbols $a^\dagger_j$ and $a_{j^\prime}$ denote creation and destruction
operators for bosons at lattice site $j$.  These operators obey the commutation relation  $[a_{j^\prime},a^\dagger_j]=\delta_{j^\prime j}$; $\hat{n}_j=a^\dagger_j a_j$ is the boson number operator at site $j$. $V_T(\mathbf{r}_j)$ is the trap potential (which is usually assumed to be a simple harmonic oscillator potential) and the repulsive contact interaction is given by $U$. Note that the trapping potential could also represent a diagonal disorder potential, if desired, but we will not discuss that case further here. The chemical potential $\mu$ controls the average number of particles. $\mathfrak{t}_{j j^\prime}$ is the amplitude for bosons to hop from site $j^\prime$ to site $j$. We consider a general $\mathfrak{t}_{j j^\prime}$ for the formal developments we present in the earlier parts of the paper, but later specialize to the case of nearest-neighbor hopping only, with amplitude $\mathfrak{t}$, on a hypercubic lattice in $d$ dimensions.

As explained above, our aim in this paper is to calculate the properties of the Hamiltonian in
Eq.~(\ref{eq: hubham}), in particular, its momentum distribution function.
The momentum distribution function is related to the  atom-atom correlation function [see Eq.~(\ref{eq: nk_def})] involving atoms at sites $j$ and $j^\prime$, which is given by
\beq
C_{j^\prime j} = \langle a^{\dagger}_{j^\prime} a_j \rangle_\mathcal{H}; \; \;
\langle \mathcal{A} \rangle_\mathcal{H}  \equiv  Z^{-1}{\rm Tr}[\mathcal{A}e^{-\beta \mathcal{H}}].
\eeq
This expectation value can be calculated from the single-particle ``thermal'' or ``Matsubara''  Green's function, defined in the standard way~\cite{mahan}, as
\beq
G_{j j^\prime}(\tau, \tau^\prime) \equiv  - \langle \mathcal{T}_{\tau} [ e^{\tau \mathcal{H}} a_j e^{-\tau \mathcal{H}} e^{\tau^\prime \mathcal{H}} a^{\dagger}_{j^\prime} e^{-\tau^\prime \mathcal{H}}]  \rangle_\mathcal{H},
\label{eq: GF-def}
\eeq
by choosing  $\tau = 0$ and $\tau^\prime = 0^+$, the positive infinitesimal; {\it i.~e.},
\beq
C_{j^\prime j} = - G_{j j^\prime}(0, 0^+).
\label{eq: CfromG}
\eeq
Here, as usual,  $\beta \equiv 1 /(k_B T)$ is the inverse temperature, $ 0 < \tau , \tau^\prime < \beta$ are ``imaginary time'' (henceforth ``{\em i-time}'') variables, $\mathcal{T}_{\tau}$ is the i-time-ordering operator,
and $ Z \equiv {\rm Tr}[e^{-\beta \mathcal{H}}]$ is the partition function. The momentum distribution function measured in the time of flight experiments is proportional to the Fourier transform of the atom-atom correlation function:
\beq
 n_{\bf k} \equiv \frac{1}{\mathcal{N}} \sum_{j, j^\prime}C_{j^\prime j} e^{  i \bf{k} \cdot (\mathbf{r}_{j^\prime}- \mathbf{r}_j)}
\label{eq: nk_def}
\eeq
where $\mathcal{N}$ is the number of sites in the lattice (we do not discuss the proportionality factors, which arise from the Wannier wavefunctions of the trapped atoms, as that is not germane to the work we present here).

Specializing to the case of nearest-neighbor hopping on a hypercubic lattice in $d$-dimensions, we report our main result which is the general strong-coupling expansion for the ($T=0$) momentum distribution of the Mott phase with a density $n_{}$ up to third order in the hopping:
\begin{eqnarray}
n_{\bf k} & = & n_{} \left \{ 1-2(n_{}+1)\frac{\epsilon_{\bf k}}{U} + 3(n_{}+1)(2n_{}+1) \left[ \left (\frac{\epsilon_{\bf k}}{U} \right )^2 - 2d \left (\frac{\mathfrak{t}}{U} \right)^2 \right ] \right . \non \\
& - & 4(n_{} + 1)[5n_{}^2+5n_{}+1] \left (\frac{\epsilon_{\bf k}}{U} \right )^3   \nonumber\\
& + & \left [\frac{2}{3}(n_{} + 1)(26 n_{}^2 + 26 n_{} + 5) \right ] 4d \left( \frac{\epsilon_{\bf k}\mathfrak{t}^2}{U^3} \right ) \nonumber\\
& - & \left . \left [\frac{1}{3}(n_{} + 1)(23 n_{}^2 + 23 n_{} + 2) \right] \left (\frac{\epsilon_{\bf k}\mathfrak{t}^2}{U^3}\right ) \right \},
\label{eq: third_order_general}
\end{eqnarray}
where $\epsilon_{\mathbf k}=-\mathfrak{t}\sum_\delta\exp[i{\mathbf k}\cdot \bf{\delta}]$ is the bandstructure, $\bf{\delta}$ is a nearest-neighbor translation vector, and $d$ is the spatial dimension.

Readers who are mainly interested in seeing how accurate this expansion is when applied to explicit cases, are encouraged to skip the next section which develops the formal techniques needed for obtaining the expansion, and proceed directly to Sec.~III where we use the expansion to develop a scaling analysis and compare results to exact numerics.

The manuscript is organized as follows:  in Sec. II, we present the formalism for the strong-coupling expansion of the Green's functions and produce explicit formulas through third order for the one-dimensional lattice, the two-dimensional square lattice and the three-dimensional cubic lattice, along with the infinite-dimensional hypercubic lattice. In Sec. III, we present our scaling analysis for the momentum distribution in the first Mott lobe and compare those results to available numerical data from QMC and DMRG calculations; we also discuss the phenomenological approach to the phase diagram in two and three dimensions. Conclusions and a discussion of future directions follow in Sec. IV. Two appendices contain some of the more technical results.

\section{Strong-Coupling Formalism for the Green's functions}

The strong-coupling expansion we develop in this paper enables one to calculate $G_{j j^\prime}$ [in Eq.~(\ref{eq: GF-def})] and hence $C_{j^\prime j}$ [in Eq.~(\ref{eq: CfromG})] as an expansion in powers of  $\mathcal{H}_{hop}$ [in Eq.~(\ref{eq: hubham})], with respect to regions of the system which are either
{\em normal or Mott-insulating (i.~e., not superfluid~\cite{sf-expn})}.  For this purpose, we use the following standard relation~\cite{mahan} to define the i-time-ordered product for the evolution operator in the ``interaction picture'':
\bea
e^{-\tau \mathcal{H}}e^{\tau^\prime \mathcal{H}}  & = & e^{-\tau \mathcal{H}_0} \mathcal{U}(\tau, \tau^\prime) e^{\tau^\prime \mathcal{H}_0}; \\
\mathcal{U}(\tau, \tau^\prime) & \equiv & \mathcal{T}_{\tau} \exp {[-\int_{\tau^\prime}^\tau d\tau_1 \mathcal{H}_{hop}(\tau_1)]}
\eea
where, for any operator $\mathcal{A}$, we define the time-dependent operator $\mathcal{A}(\tau_1) \equiv   e^{\tau_1 \mathcal{H}_0} \mathcal{A} e^{-\tau_1 \mathcal{H}_0}$. Using the properties of $\mathcal{T}_{\tau}$,  and the rules for composition for products of $\mathcal{U}$, it is straightforward to show that~\cite{mahan}
\beq
G_{jj^\prime}(\tau, \tau^\prime) = - \frac {\langle \mathcal{T}_{\tau} [\mathcal{U}(\beta, 0) a_j(\tau) a^{\dagger}_{j^\prime}(\tau^\prime)]  \rangle_{\mathcal{H}_0}} {\langle  \mathcal{U}(\beta, 0) \rangle_{\mathcal{H}_0}}
\label{eq: GF-basic}
\eeq
The strong coupling expansion we use in this paper is obtained straightforwardly by expanding the exponentials in $\mathcal{U}$ [in Eq.~(\ref{eq: GF-basic})] in powers of $\mathcal{H}_{hop}$ and evaluating the resulting traces with respect to the equilibrium ensemble of $\mathcal{H}_0$. The term of order $m$ in such an expansion for the {\em numerator} in Eq.~(\ref{eq: GF-basic}) is given by
\bea
\frac{1}{m!}  \sum_{j_m  j^\prime_m} \cdots  \sum_{j_1 j^\prime_1}  \int_0^\beta d {\tau}_m  \cdots \int_0^\beta d {\tau}_1 \; \mathfrak{t}_{j_m j^\prime_m} \cdots \mathfrak{t}_{j_1 j^\prime_1}  \non\\ \times \langle \mathcal{T}_{\tau}[ a_j(\tau)   a^{\dagger}_{j_m}({\tau}_m^+) a_{j^\prime_m}({\tau}_m)  \cdots a^{\dagger}_{j_1}({\tau}_1^+) a_{j^\prime_1}({\tau}_1) a^{\dagger}_{j^\prime}(\tau^\prime)] \rangle_{\mathcal{H}_0}.
\label{eq: Nr-GF-m}
\eea
Since $\mathcal{H}_0$, as defined in Eq.~(\ref{eq: hubham}), is a sum of separate terms for each site, the thermal average in Eq.~(\ref{eq: Nr-GF-m}) factorizes into a product of factors, one for each of the sites on the lattice, in terms of the multiparticle single-site Green's functions at these sites defined in the standard way~\cite{mahan} as,
\bea
\mathcal{G}_{j}(\tau_1,\tau^\prime_1) &\equiv& -  \langle \mathcal{T}_{\tau}[ a_j(\tau_1) a^{\dagger}_{j}(\tau^\prime_1)]  \rangle_{\mathcal{H}_{0 j}} \label{eq: calgI-def}\\
\mathcal{G}_{j}^{II}(\tau_1,\tau_2;\tau_2^\prime,\tau_1^\prime) &\equiv&  \langle \mathcal{T}_{\tau}[ a_j(\tau_1) a_j(\tau_2) a^{\dagger}_{j}(\tau_2^\prime) a^{\dagger}_{j}(\tau_1^\prime)] \rangle_{\mathcal{H}_{0 j}}
\label{eq: calgII-def}\\
\vdots \non
\eea
Note that these are total Green's functions, containing both connected and disconnected parts.
Furthermore, each site that appears must occur an even number of times in the thermal average, half as indices of creation operators and half as indices of destruction operators~\cite{sf-expn}. Similar considerations apply to the terms in the expansion for the denominator in Eq.~(\ref{eq: GF-basic}) [except for the absence of the operators $a_j(\tau)$ and $a^{\dagger}_{j^\prime}(0)$]. As we discuss in more detail below, the combination of the two expansions order by order leads to a cancelation of all ``disconnected'' terms, i.e., those involving products of thermal averages for clusters of sites that are not connected via hopping matrix elements to the sites $j$ and $j^\prime$, as well as to the fact that the remaining terms can be written entirely in terms of the ``connected'' or ``cumulant'' multiparticle Greens functions, corresponding to the well known linked cluster theorem~\cite{mahan}.

\begin{figure}[th]
\centerline{\includegraphics [width=4.3in, angle=0, clip=on]  {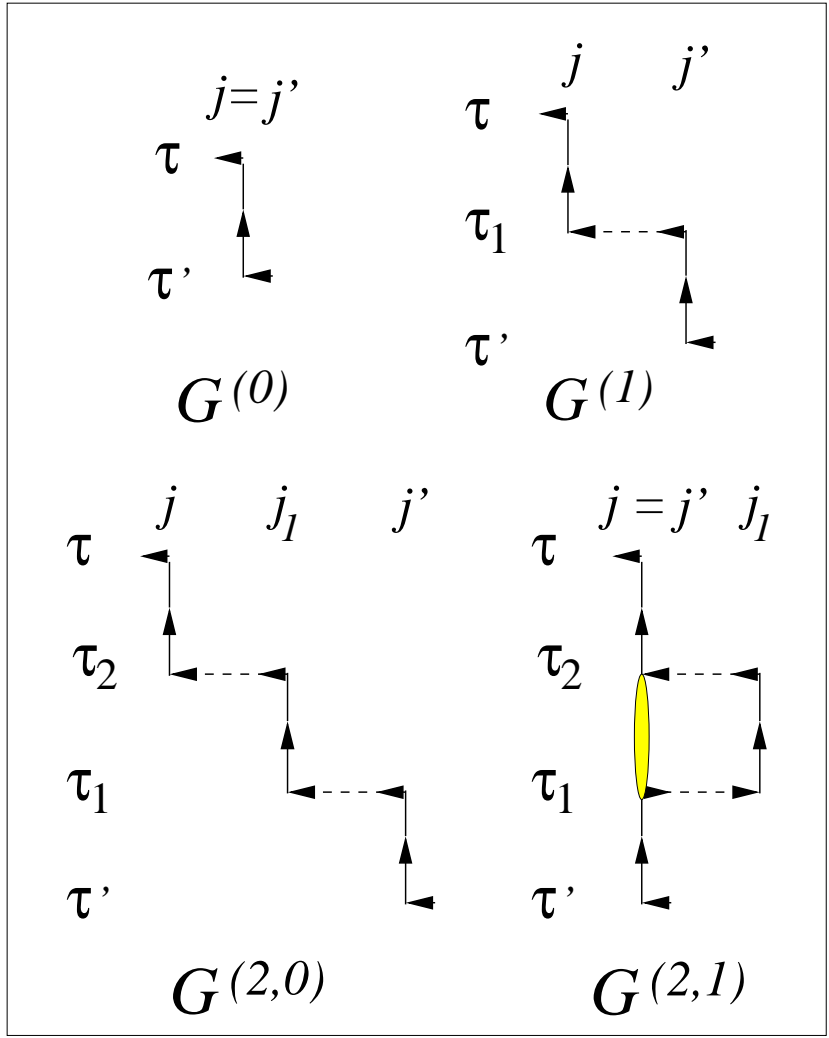}}
\caption[]{(Color online.) Strong-coupling ``diagrams'' for the single-particle Green's functions up to second order in $\mathfrak{t}$. The horizontal directed dashed lines indicate the hopping matrix element $\mathfrak{t}$ between the sites labeled, and the vertical lines indicate single-site Green's functions $\mathcal{G}$ evolving between the respective i-times. The ellipses (yellow) at multiply visited sites denote the appearance of connected or cumulant $n$-particle Green's functions.}
\label{fig: G-diag-1}
\end{figure}

\begin{figure}[th]
\centerline{\includegraphics [width=4.3in, angle=0, clip=on]  {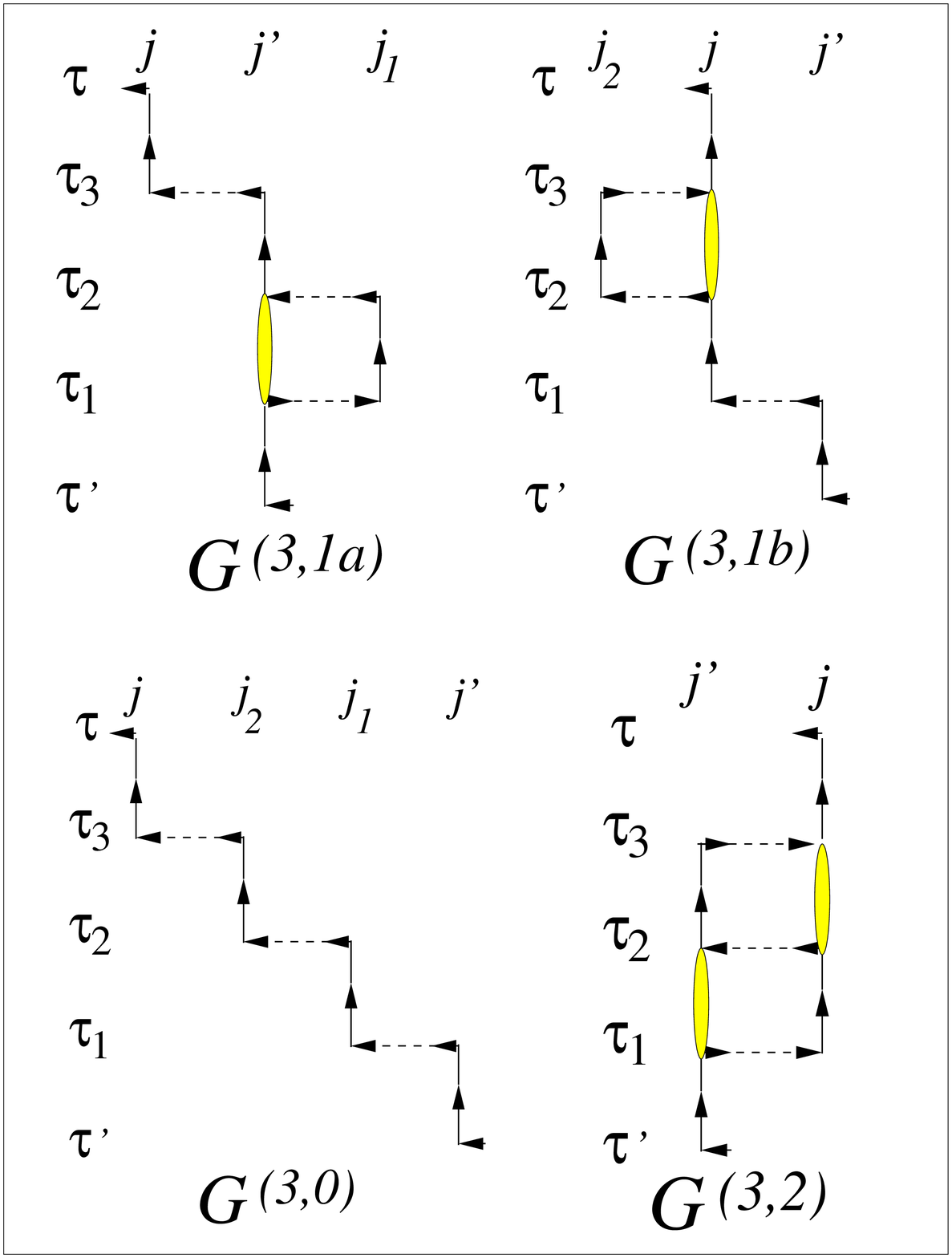}}
\caption[]{(Color online.) Strong-coupling ``diagrams'' for the single-particle Green's functions corresponding to third order in $\mathfrak{t}$. The horizontal directed dashed lines indicate the hopping matrix element $\mathfrak{t}$ between the sites labeled, and the vertical lines indicate single-site Green's functions $\mathcal{G}$ evolving between the respective i-time. The ellipses (yellow) at multiply visited sites denote the appearance of connected or cumulant $n$-particle Green's functions.}
\label{fig: G-diag-2}
\end{figure}

Using the above considerations, it is straightforward to write down systematically the terms in the strong-coupling expansion for $G_{jj^\prime}(\tau, \tau^\prime)$. We denote the $m^{th}$ order contributions with a superscript $(m)$. The different terms contributing in $m^{th}$ order can also be associated with ``diagrams'', which correspond to lattice ``walks'' or ``world lines'' for a particle which starts from site $j^\prime$ at i-time $\tau^\prime$ and reaches site $j$ at i-time $\tau$ after $m$ steps (with each `step' corresponding to a hop along the lattice, {\it e.~g.}, from site $j^\prime_1$ to site $j_1$  induced by $\mathfrak{t}_{j_1 j^\prime_1}$, and in between the steps, the particle undergoes i-time ``evolution'', which proceeds either forward or backward in i-time). These processes are shown graphically in Figs. \ref{fig: G-diag-1} and \ref{fig: G-diag-2}. These diagrams are the strong-coupling analogs of the standard diagrams of many-body perturbation theory~\cite{mahan}, from which, after some practice, the terms can be written down by inspection. A $p$ particle Green's function at a particular site appears when a walk visits that site $p$ times. For $m \ge 2$, as we show below, the contributions can be classified further according to a hierarchy of decreasing powers of $1/z$, where $z$ is the coordination number of the lattice, by recombining contributions from intersecting and nonintersecting walks, and we denote these with further superscripts, as $(m;0),(m;1),$ {\it etc}.  We give below the terms contributing to $G_{jj^\prime}(\tau, \tau^\prime)$ up to third order in  $\mathcal{H}_{hop}$, and their associated strong-coupling diagrams.

The zeroth and first order terms are almost obvious.
\beq
G_{jj^\prime}^{(0)}(\tau, \tau^\prime) =  \delta_{j j^\prime} \mathcal{G}_{j}(\tau, \tau^\prime),
\label{eq: G0}
\eeq
\beq
G_{jj^\prime}^{(1)}(\tau, \tau^\prime)  =  - \mathfrak{t}_{j j^\prime} \int_{\tau_1} \mathcal{G}_{j}(\tau,\tau_1) \mathcal{G}_{j^\prime}(\tau_1, \tau^\prime) \equiv - \mathfrak{G}^{(1)}_{j j^\prime} (\tau, \tau^\prime).
\label{eq: G1}
\eeq
Here, and below, for notational convenience we denote integrals over i-times by integral symbols with subscripts, rather than by the standard notation.
To second order, a 2-step lattice walk can either move to a distinct site two steps away or return to the starting site. Hence we get two terms from the numerator of Eq.~(\ref{eq: GF-basic}), the top equation is when the hop is to a different lattice site, the bottom equation is when the hop returns back to the original lattice site:
\bea
G_{jj^\prime}^{(2;a)}(\tau, \tau^\prime)_{num} & = & (1- \delta_{j j^\prime}) \sum_{j_1} \mathfrak{t}_{j j_1} \mathfrak{t}_{j_1 j^\prime} \int_{\tau_2} \int_{\tau_1} \mathcal{G}_{j}(\tau,\tau_2) \mathcal{G}_{j_1}(\tau_2,\tau_1) \mathcal{G}_{j^\prime}(\tau_1, \tau^\prime), \\
& \equiv & (1- \delta_{j j^\prime}) \, \sum_{j_1} \mathfrak{G}^{(2)}_{j j_1 j^\prime} (\tau, \tau^\prime)\\
G_{jj^\prime}^{(2;b)}(\tau, \tau^\prime)_{num} & = & \delta_{j j^\prime} \sum_{j_1} \mathfrak{t}_{j j_1} \mathfrak{t}_{j_1 j} \int_{\tau_2} \int_{\tau_1} \mathcal{G}_{j}^{II}(\tau,\tau_1;\tau_2, \tau^\prime) \mathcal{G}_{j_1}(\tau_2,\tau_1) \\
& \equiv &  \delta_{j j^\prime}  \sum_{j_1} \bar{\mathfrak{G}}^{(2)}_{j j_1 j} (\tau, \tau^\prime),
\eea
where the subscript {\it num} denotes that these are the terms coming from the numerator in the expansion for the Green's function.

{\em We have introduced a new notation above}, letting  $\mathfrak{G}_{j \cdots j^\prime}^{(m)}$ denote the product of {\em single-particle Green's functions} at the sites that appear in the $m$-step lattice walk specified by its lattice indices, starting from right to left, together with the corresponding hopping amplitudes; the i-time arguments indicating the starting and ending i-time, the $m$ intermediate i-times being integrated over. $\bar{\mathfrak{G}}_{j \cdots j^\prime}^{(m)}$ is defined similarly, except that it necessarily involves self-intersecting lattice walks where one or more sites are visited multiple times, and the product involves $r$-particle Green's functions at a site that is visited $r$ times, with the intermediate i-time arguments being determined by the sequence specified in the lattice walk. For any given $m$-step lattice walk both $\mathfrak{G}^{(m)}$ and $\bar{\mathfrak{G}}^{(m)}$ can clearly be written down by inspection.

To correctly obtain obtain $G_{jj^\prime}^{(2)}$, we need to subtract from the above two terms the term that arises as the product of the second-order contribution from the denominator of Eq.~(\ref{eq: GF-basic}), corresponding to closed loop lattice walks involving the sites $j$ and $j_1$, given by
\beq
\mathcal{Z}^{(2)}_{j j_1} \equiv  \mathfrak{t}_{j j_1} \mathfrak{t}_{j_1 j}  \int_{\tau_2} \int_{\tau_1} \mathcal{G}_{j}(\tau_1, \tau_2) \mathcal{G}_{j_1}(\tau_2,\tau_1),
\eeq
and the zeroth-order term from the numerator, namely $G_{jj^\prime}^{(0)}(\tau, \tau^\prime)$.
The net result for $G^{(2)}$ can be reexpressed as the sum of the following two contributions:
\bea
G_{jj^\prime}^{(2;0)}(\tau, \tau^\prime) & = &  \sum_{j_1} \mathfrak{G}^{(2)}_{j j_1 j^\prime} (\tau, \tau^\prime)
\label{eq: G2-0} \\
G_{jj^\prime}^{(2;1)}(\tau, \tau^\prime) & = & \delta_{j j^\prime} \sum_{j_1} \mathfrak{t}_{j j_1} \mathfrak{t}_{j_1 j} \int_{\tau_2} \int_{\tau_1} \tilde{\mathcal{G}}_{j}^{II}(\tau,\tau_1;\tau_2, \tau^\prime) \mathcal{G}_{j_1}(\tau_2,\tau_1) \non \\
& \equiv & \delta_{j j^\prime} \sum_{j_1} \tilde{\mathfrak{G}}^{(2)}_{j j_1 j} (\tau, \tau^\prime) \\
& = & \delta_{j j^\prime} \sum_{j_1} \{  \bar{\mathfrak{G}}^{(2)}_{j j_1 j} (\tau, \tau^\prime) - \mathfrak{G}^{(2)}_{j j_1 j} (\tau, \tau^\prime) -  \mathcal{G}_{j}(\tau, \tau^\prime) \mathcal{Z}^{(2)}_{j j_1} \}.
\label{eq: G2-1}
\eea
Here
\beq
\tilde{\mathcal{G}}_{j}^{II}(\tau,\tau_1;\tau_2, \tau^\prime) \equiv [\mathcal{G}_{j}^{II}(\tau,\tau_1;\tau_2, \tau^\prime) - \mathcal{G}_{j}(\tau,\tau_2)\mathcal{G}_{j}(\tau_1, \tau^\prime) - \mathcal{G}_{j}(\tau, \tau^\prime)\mathcal{G}_{j}(\tau_1, \tau_2)]
\label{eq: calgII-conn}
\eeq
is the ``cumulant'' or ``connected'' part of the two-particle Green's function at site $j$. $\tilde{\mathfrak{G}}^{(m)}$ is defined  similarly to $\bar{\mathfrak{G}}^{(m)}$ {\em except that the multi-particle Green's functions that appear in $\tilde{\mathfrak{G}}^{(m)}$  are all connected Green's functions}. Note that the prefactor $\frac{1}{2!}$ present in Eq.~(\ref{eq: Nr-GF-m}) no longer appears in the above equations, as it has been canceled by the $2!$ ways of choosing the two distinct hopping matrix elements in the expansion.

Similarly, the third-order contributions involve three-step walks. From the numerator of Eq.~(\ref{eq: GF-basic}) we get the following terms according to the types of walks involved.
\bea
G_{jj^\prime}^{(3;a)}(\tau, \tau^\prime)_{num} & = & - (1- \delta_{j j^\prime})\sum_{j_2, j_1}(1- \delta_{j_2 j^\prime})(1- \delta_{j j_1}) \; \mathfrak{t}_{jj_2} \mathfrak{t}_{j_2 j_1} \mathfrak{t}_{j_1 j^\prime} \non \\ &\times& \int_{\tau_3} \int_{\tau_2} \int_{\tau_1}  \mathcal{G}_{j}(\tau, \tau_3)\mathcal{G}_{j_2}(\tau_3, \tau_2) \mathcal{G}_{j_1}(\tau_2, \tau_1) \mathcal{G}_{j^\prime}(\tau_1, \tau^\prime),\\
G_{jj^\prime}^{(3;b)}(\tau, \tau^\prime)_{num} & = & - (1- \delta_{j j^\prime})\sum_{j_1}  (1- \delta_{j j_1}) \; \mathfrak{t}_{j j^\prime} \mathfrak{t}_{j^\prime j_1} \mathfrak{t}_{j_1 j^\prime}  \non \\ &\times& \int_{\tau_3} \int_{\tau_2} \int_{\tau_1}  \mathcal{G}_{j}(\tau, \tau_3) \mathcal{G}_{j_1}(\tau_2, \tau_1) \mathcal{G}^{II}_{j^\prime}(\tau_1, \tau_3; \tau_2, \tau^\prime) \non \\
&& - (1- \delta_{j j^\prime})\sum_{j_2}(1- \delta_{j_2 j^\prime}) \; \mathfrak{t}_{j j_2} \mathfrak{t}_{j_2 j} \mathfrak{t}_{j j^\prime} \non \\ &\times& \int_{\tau_3} \int_{\tau_2} \int_{\tau_1}  \mathcal{G}^{II}_{j}(\tau, \tau_2; \tau_3, \tau_1) \mathcal{G}_{j_2}(\tau_3, \tau_2)  \mathcal{G}_{j^\prime}(\tau_1, \tau^\prime),\\
G_{jj^\prime}^{(3;c)}(\tau, \tau^\prime)_{num} & = & - \frac{1}{2!}(1- \delta_{j j^\prime}) \; \mathfrak{t}_{j j^\prime} \mathfrak{t}_{j^\prime j} \mathfrak{t}_{j j^\prime} \non \\ &\times& \int_{\tau_3} \int_{\tau_2} \int_{\tau_1}  \mathcal{G}^{II}_{j}(\tau, \tau_2; \tau_3, \tau_1) \mathcal{G}^{II}_{j^\prime}(\tau_3, \tau_1; \tau_2, \tau^\prime).
\eea
Again, in all cases except for the case of $G^{(3;c)}_{num}$, the $3!$ ways of choosing the three distinct hopping matrix elements involved completely cancels the $\frac{1}{3!}$ in the expansion. In case of $G^{(3;c)}_{num}$, two of the hopping matrix elements are identical, so they can be chosen in only $\frac{3!}{2!}$ ways, hence there is a factor of $\frac{1}{2!}$ left uncanceled.

The restriction $(1- \delta_{j j^\prime})$ in the third-order contributions above is redundant except on nonbipartite lattices, such as nearest-neighbor hopping on a triangular lattice, or on a hypercubic lattice with second-neighbor hopping, where one can return to the starting site after three hops. In such cases, one has the additional term
\bea
G_{jj^\prime}^{(3;d)}(\tau, \tau^\prime)_{num} & = & - \delta_{j j^\prime}\sum_{j_2, j_1}(1- \delta_{j_2 j})(1- \delta_{j j_1}) \; \mathfrak{t}_{jj_2} \mathfrak{t}_{j_2 j_1} \mathfrak{t}_{j_1 j} \non \\ &\times& \int_{\tau_3} \int_{\tau_2} \int_{\tau_1}  \mathcal{G}^{II}_{j}(\tau, \tau_1; \tau_3, \tau^\prime)\mathcal{G}_{j_2}(\tau_3, \tau_2) \mathcal{G}_{j_1}(\tau_2, \tau_1).
\eea
Note that in this term the constraints on $j_1$ and $j_2$ are actually redundant and can be omitted.

As in the second-order case, the above third-order terms can be recombined with the terms that arise as products of the first-order term from the numerator and the appropriate second-order terms from the denominator in Eq.~(\ref{eq: GF-basic}) and reexpressed compactly in terms of the connected Green's functions $\mathfrak{G}^{(3)}$ and $\tilde{\mathfrak{G}}^{(3)}$. One gets
\bea
G_{jj^\prime}^{(3;0)}(\tau, \tau^\prime) & = & - \sum_{j_2, j_1} \mathfrak{G}^{(3)}_{j j_2 j_1 j^\prime} (\tau, \tau^\prime),
\label{eq: G3-0} \\
G_{jj^\prime}^{(3;1)}(\tau, \tau^\prime) & = & -  \sum_{j_1} \tilde{\mathfrak{G}}^{(3)}_{j j^\prime j_1 j^\prime} (\tau, \tau^\prime) - \sum_{j_2} \tilde{\mathfrak{G}}^{(3)}_{j j_2 j j^\prime} (\tau, \tau^\prime),
\label{eq: G3-1} \\
G_{jj^\prime}^{(3;2)}(\tau, \tau^\prime) & = & -  \frac{1}{2!} \tilde{\mathfrak{G}}^{(3)}_{j j^\prime j j^\prime} (\tau, \tau^\prime).
\label{eq: G3-2}
\eea
In  nonbipartite cases, one has to add to this the additional contribution
\beq
G_{jj^\prime}^{(3;3)}(\tau, \tau^\prime)  =  - \delta_{j j^\prime} \sum_{j_2, j_1} \tilde{\mathfrak{G}}^{(3)}_{j j_2 j_1 j} (\tau, \tau^\prime).
\label{eq: G3-3}
\eeq
Note that in all the cases, use of the connected Green's functions allows one to avoid the clumsy restrictions on the intermediate sites that need to be summed over, and in addition, automatically includes the terms contributed by the denominator of Eq.~(\ref{eq: GF-basic}). The diagrams that represent the above are shown in Figs.~\ref{fig: G-diag-1} and \ref{fig: G-diag-2}. The same results can also be derived using more formal methods, such as functional integrals, generating functionals, and functional derivatives, but we do not go into such details here.

Next, we discuss the evaluation of the multiparticle single-site Green's functions at a site $j$ as defined in Eqs.~(\ref{eq: calgI-def}) and (\ref{eq: calgII-def}). The eigenstates of $\mathcal{H}_{0j}$ in Eq.~(\ref{eq: hubham}) are also eigenstates of the number operator $\hat{n}_j$, and can hence be labeled by positive integers $n = 0, 1, \cdots$ corresponding to the number of bosons at site $j$, with energy eigenvalues which we label as $\epsilon_{j, n}$. One has,
\beq
\mathcal{H}_{0j} |j, n \rangle  =  \epsilon_{j, n} |j, n \rangle ;  \;
\epsilon_{j, n}  \equiv  [V_T (\mathbf{r}_j)\ -\mu]{n} + \frac{U}{2} {n}({n}-1).
\eeq
The partition function of the $j^{th}$ site, and the Boltzmann probability of occupancy of  $|j, n \rangle$ in the thermal ensemble corresponding to $\mathcal{H}_{0j}$, are given respectively by
\beq
\mathcal{Z}_j = \sum_{n} \exp{(-\beta \epsilon_{j, n})} ; \; \rho_{j, n} =  \exp{(-\beta \epsilon_{j, n})} / \mathcal{Z}_j .
\eeq
It is convenient to define the ladder operators
\beq
\mathcal{X}^+_{j, n} \equiv  |j, n + 1 \rangle \langle j, n | , \;  \mathcal{X}^-_{j, n} \equiv  |j, n - 1 \rangle \langle j, n | .
\eeq
One can easily see that
\beq
a_j(\tau) =  \sum_{n} e^{\tau \epsilon^-_{j, n}} \sqrt{n} \, \mathcal{X}^-_{j, n} , \; a^{\dagger}_{j}(\tau) = \sum_{n} e^{\tau \epsilon^+_{j, n}} \sqrt{n + 1} \, \mathcal{X}^+_{j, n} ,
\eeq
where
\beq
\epsilon^+_{j, n} \equiv \epsilon_{j, (n + 1)} - \epsilon_{j, n} , \; \epsilon^-_{j, n} \equiv \epsilon_{j, (n - 1)} - \epsilon_{j, n}
\eeq
are the ``particle'' and ``hole'' ``excitation energies'' (with respect to the state with $n$ bosons at site $j$)  induced by the ladder operators $\mathcal{X}^+_{j, n_j}$ and $\mathcal{X}^-_{j, n_j}$, respectively.

Using the above, and the rather obvious rules for products of the ladder operators, it is easy to verify that, for the 1-particle Green's function, we have
\beq
\mathcal{G}_{j}(\tau_1, \tau_2) = \sum_{n} \rho_{j, n} [ (n +1) \, e^{(\tau_2 - \tau_1) \epsilon^+_{j, n}} \, \theta (\tau_1 - \tau_2) +  n \, e^{(\tau_1 - \tau_2) \epsilon^-_{j, n}} \, \theta (\tau_2 -\tau_1 ) ].
\label{eq: calgI}
\eeq
There is a compact way of working and writing this out which easily generalizes to $n$-particle Green's functions. Let $\mathbb{P}$ label the $2!$ possible permutations of $(1,2)$, corresponding to $(1,2) \rightarrow (\mathbb{P}1, \mathbb{P}2)$, and  $\mathcal{G}_{j}(\mathbb{P})$ denote $\mathcal{G}_{j}(\tau_1, \tau_2)$ in the domain $(\tau_{\mathbb{P}1} > \tau_{\mathbb{P}2})$. Furthermore, define $\sigma_{1} \equiv -1$ , $\sigma_{2} \equiv +1$ ; $\epsilon^{\pm 1}_{j, n} \equiv  \epsilon^{\pm}_{j, n}$; and  $\mathcal{X}^{\pm 1}_{j, n} \equiv  \mathcal{X}^{\pm}_{j, n}$. Then, one has,
\bea
\mathcal{G}_j(\mathbb{P}) & = & \sum_{n} \rho_{j, n} \sum_{n_1, n_2} \langle j, n| \left[ \prod_{\ell = 1,2} \exp{\tau_{\mathbb{P}\ell} \epsilon^{\sigma_{\mathbb{P}\ell}}_{j, n_\ell}} \, \sqrt{n_\ell + \frac{1 + \sigma_{\mathbb{P} \ell}}{2}} \, \mathcal{X}^{\sigma_{\mathbb{P}\ell}}_{j, n_\ell} \right ] \, |j, n\rangle \non \\
& = & \sum_{n} \rho_{j, n} \sqrt{n + \frac{1 - \sigma_{\mathbb{P}1}}{2}} \sqrt{n + \frac{1 + \sigma_{\mathbb{P}2}}{2}} \exp{[ \tau_{\mathbb{P}2} \epsilon^{\sigma_{\mathbb{P}2}}_{j, n} -  \tau_{\mathbb{P}1} \epsilon^{-\sigma_{\mathbb{P}1}}_{j, n}]}.
\label{eq: calgI-P}
\eea
As is easily verified, for the identity permutation, corresponding to $\mathbb{P}1 = 1, \mathbb{P}2 =2 $, this reproduces the first term in Eq.~(\ref{eq: calgI}); for the permutation corresponding to $\mathbb{P}1 = 2, \mathbb{P}2 = 1$, it reproduces the second term in Eq.~(\ref{eq: calgI}).

Now, for the case of the 2-particle Green's functions, let $\mathbb{P}$ label the $4!$ possible permutations of $(1, 2, 3, 4)$, corresponding to $(1, 2, 3, 4) \rightarrow (\mathbb{P}1, \mathbb{P}2, \mathbb{P}3, \mathbb{P}4)$, and  $\mathcal{G}^{II}_{j}(\mathbb{P})$ denote $\mathcal{G}^{II}_{j}(\tau_1, \tau_2 ; \tau_3, \tau_4)$ in the domain $(\tau_{\mathbb{P}1} > \tau_{\mathbb{P}2} > \tau_{\mathbb{P}3} > \tau_{\mathbb{P}4})$. For this case, we define $\sigma_{1} = \sigma_{2} \equiv -1$  and $\sigma_{3} = \sigma_{4} \equiv +1$. Using these definitions, we can show that
\bea
\mathcal{G}^{II}_{j}(\mathbb{P}) & = & \sum_{n} \rho_{j, n} \sum_{n_1, \cdots, n_4} \langle j, n| \left[ \prod_{\ell = 1,\cdots, 4} \exp{\tau_{\mathbb{P}\ell} \epsilon^{\sigma_{\mathbb{P}\ell}}_{j, n_\ell}} \, \sqrt{n_\ell + \frac{1 + \sigma_{\mathbb{P} \ell}}{2}} \, \mathcal{X}^{\sigma_{\mathbb{P}\ell}}_{j, n_\ell} \right ] \, |j, n\rangle \non \\
& = & \sum_{n} \rho_{j, n} \sqrt{n + \frac{1 - \sigma_{\mathbb{P}1}}{2}} \sqrt{n - \sigma_{\mathbb{P}1} + \frac{1 - \sigma_{\mathbb{P}2}}{2}} \sqrt{n + \sigma_{\mathbb{P}4} + \frac{1 + \sigma_{\mathbb{P}3}}{2}} \sqrt{n + \frac{1 + \sigma_{\mathbb{P}4}}{2}} \non \\
& \times & \exp{[ \tau_{\mathbb{P}4} \epsilon^{\sigma_{\mathbb{P}4}}_{j, n} + \tau_{\mathbb{P}3} \epsilon^{\sigma_{\mathbb{P}3}}_{j, (n + \sigma_{\mathbb{P}4})} -  \tau_{\mathbb{P}2} \epsilon^{-\sigma_{\mathbb{P}2}}_{j, n - \sigma_{\mathbb{P}1}} - \tau_{\mathbb{P}1} \epsilon^{-\sigma_{\mathbb{P}1}}_{j, n} ]}.
\label{eq: calgII-P}
\eea
For example, in the domain $(\tau_1 > \tau_2 > \tau_3 > \tau_4)$, corresponding to the identity permutation, this formula gives,
\bea
\mathcal{G}^{II}_{j} & = & \langle a_j(\tau_1) a_j(\tau_2) a^{\dagger}_{j}(\tau_3) a^{\dagger}_{j}(\tau_4) \rangle_{\mathcal{H}_{0 j}} \non \\
& = & \sum_{n} \rho_{j, n} (n + 1) (n + 2) \exp{[(\tau_4 - \tau_1) \epsilon^+_{j, n} + (\tau_3 - \tau_2) \epsilon^+_{j, n + 1} ]}.
\eea

Using the above results, we can now readily compute the terms in the strong-coupling expansion of $C_{j^\prime j}$ in Eq.~(\ref{eq: CfromG}) up to third order in the hopping amplitude. In the equations below, we denote
$\mathfrak{C}_{j \cdots j^\prime}^{(m)} \equiv \mathfrak{G}_{j \cdots j^\prime}^{(m)}(0,0^+)$, $\bar{\mathfrak{C}}_{j \cdots j^\prime}^{(m)} = \bar{\mathfrak{G}}_{j \cdots j^\prime}^{(m)}(0,0^+)$, and $\tilde{\mathfrak{C}}_{j \cdots j^\prime}^{(m)} \equiv \tilde{\mathfrak{G}}_{j \cdots j^\prime}^{(m)}(0,0^+)$.

\begin{figure}[th]
\centerline{\includegraphics [width=3.3in, angle=0, clip=on]  {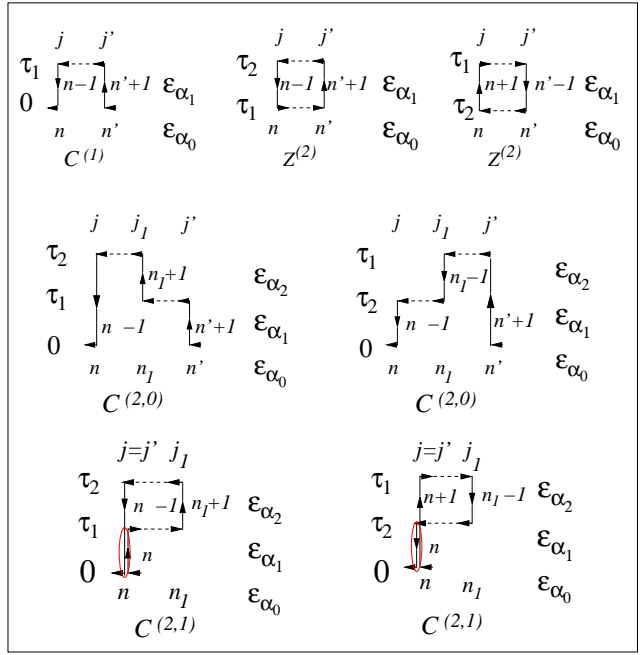}}
\caption[]{Strong coupling ``diagrams'' for the correlation functions up to second order in $\mathfrak{t}$. The horizontal directed dashed lines indicate $\mathfrak{t}$ between the sites labeled, and the vertical lines indicate single site Greens functions $\mathcal{G}$. The ellipses (red on line) at multiply visited sites denote the appearance of connected or cumulant Greens functions. Initial and intermediate state labels for the different possible i-time orderings shown are also indicated.}
\label{fig: C-diag-1}
\end{figure}

The zeroth order term is
\beq
C^{(0)}_{j^\prime j}  =  - G^{(0)}_{j j^\prime}(0, 0^+) =  \delta_{j, j^\prime} \langle \hat{n}_j \rangle_{\mathcal{H}_{0 j}} = \delta_{j, j^\prime} \sum_{n} \; n \; \rho_{j, n}.
\eeq
The first order term is
\beq
C^{(1)}_{j^\prime j} =  - G^{(1)}_{j j^\prime}(0, 0^+)  =    \mathfrak{t}_{j j^\prime}  \int_{\tau_1}  \mathcal{G}_{j}(0, \tau_1) \mathcal{G}_{j^\prime}(\tau_1, 0^+) \equiv \mathfrak{C}_{j j^\prime}^{(1)}.
\eeq
The i-time integral is straightforward to evaluate. Using Eq.~(\ref{eq: calgI}) or Eq.~(\ref{eq: calgI-P}), we find,
\bea
\mathfrak{C}_{j j^\prime}^{(1)} & = & \mathfrak{t}_{j j^\prime} \sum_{n, n^\prime}  n (n^\prime + 1) \; \rho_{j, n} \, \rho_{j^\prime, n^\prime}    \int_0^{\beta} d\tau_1 \exp{[-\tau_1 (\epsilon^-_{j, n} +  \epsilon^+_{j^\prime, n^\prime})]} \non \\
& = & \mathfrak{t}_{j j^\prime} \sum_{n, n^\prime} n (n^\prime + 1) \; \rho_{j, n} \, \rho_{j^\prime, n^\prime}  \left \{  \frac {1-\exp{ [- \beta (\epsilon^-_{j, n} +  \epsilon^+_{j^\prime, n^\prime})]}} {\epsilon^-_{j, n} +  \epsilon^+_{j^\prime, n^\prime}} \right \}  \non \\
& = &  \mathfrak{t}_{j j^\prime} \sum_{n, n^\prime} n (n^\prime + 1) \left  \{ \frac{\rho_{j, n} \, \rho_{j^\prime, n^\prime}}{(\epsilon^-_{j, n} +  \epsilon^+_{j^\prime, n^\prime})} + \frac{\rho_{j, n-1} \, \rho_{j^\prime, n^\prime + 1}}{(\epsilon^+_{j, n-1} +  \epsilon^-_{j^\prime, n^\prime + 1})} \right \} \non \\
& = &  \mathfrak{t}_{j j^\prime} \sum_{n, n^\prime} \rho_{j, n} \, \rho_{j^\prime, n^\prime} \left  \{ \frac{n (n^\prime + 1)}{(\epsilon^-_{j, n} +  \epsilon^+_{j^\prime, n^\prime})} + \frac{(n + 1) n^\prime}{(\epsilon^+_{j, n} +  \epsilon^-_{j^\prime, n^\prime})} \right \}.
\label{eq: C1}
\eea
This can be represented by the diagram labeled $C^{(1)}$ in Fig.~\ref{fig: C-diag-1}. The third line of Eq.~(\ref{eq: C1}) is written in a form that can be directly constructed from this diagram in a way that is immediately generalizable to higher order (see below). The fourth line contains a second form of the same result, obtained by  relabeling the bosonic occupation numbers in the second term of the third line in a way that makes the zero temperature limit obvious.

There are two second order terms in $C_{j^\prime j}$ corresponding to the two terms in $G_{j j^\prime}$ [Eqs.~(\ref{eq: G2-0}) and (\ref{eq: G2-1})].
\beq
C^{(2,0)}_{j^\prime j}  =  - G_{j j^\prime}^{(2;0)}(0, 0^+)  = - \sum_{j_1} \mathfrak{C}_{j j_1 j^\prime}^{(2)},
\label{eq: C2-0-1}\\
\eeq
and
\bea
C^{(2,1)}_{j^\prime j} & = & - G_{jj^\prime}^{(2;1)}(0, 0^+)  =  - \delta_{j j^\prime} \sum_{j_1} \tilde{\mathfrak{C}}_{j j_1 j}^{(2)} \non \\
& = &  - \delta_{j j^\prime} \sum_{j_1} \{ \bar{\mathfrak{C}}_{j j_1 j}^{(2)} - [\mathfrak{C}_{j j_1 j^\prime}^{(2)}]_{j^\prime = j} - \langle \hat{n}_j \rangle_{\mathcal{H}_{0j}} \mathcal{Z}^{(2)}_{j j_1} \}.
\label{eq: C2-1-1}
\eea
Similarly, one obtains the various terms contributing to $C_{j^\prime j}$ in third order, which we label $C^{(3,0)}_{j^\prime j}$, $C^{(3,1)}_{j^\prime j}$, $\cdots$, by setting $\tau=0$ and $\tau^\prime = 0^+$ in Eqs.~(\ref{eq: G3-0}), (\ref{eq: G3-1}), $\cdots$. One finds
\bea
C^{(3,0)}_{j^\prime j} & = &  \sum_{j_2, j_1} \mathfrak{C}^{(3)}_{j j_2 j_1 j^\prime},
\label{eq: C3-0} \\
C^{(3,1)}_{j^\prime j} & = &  \sum_{j_1} \tilde{\mathfrak{C}}^{(3)}_{j j^\prime j_1 j^\prime}  + \sum_{j_2} \tilde{\mathfrak{G}}^{(3)}_{j j_2 j j^\prime} \non \\
& = & \sum_{j_1} \{ \bar{\mathfrak{C}}_{j j^\prime j_1 j^\prime}^{(3)} - \mathfrak{C}_{j j^\prime j_1 j^\prime}^{(3)} - \mathfrak{C}_{j j^\prime}^{(1)} \mathcal{Z}^{(2)}_{j^\prime j_1} \} \non \\
&& + \sum_{j_2} \{ \bar{\mathfrak{C}}_{j j_2 j j^\prime}^{(3)} - \mathfrak{C}_{j j_2 j j^\prime}^{(3)} - \mathfrak{C}_{j j^\prime}^{(1)} \mathcal{Z}^{(2)}_{j_2 j} \},
\label{eq: C3-1} \\
C^{(3,2)}_{j^\prime j} & = &   \frac{1}{2!} \tilde{\mathfrak{C}}^{(3)}_{j j^\prime j j^\prime} \non \\
& = & \frac{1}{2!} \{ \bar{\mathfrak{C}}^{(3)}_{j j^\prime j j^\prime} - 2 [\bar{\mathfrak{C}}_{j j^\prime j_1 j^\prime}^{(3)}]_{j_1 = j} - 2 [\bar{\mathfrak{C}}_{j j_2 j j^\prime}^{(3)}]_{j_2 = j^\prime} + 2 \mathfrak{C}^{(3)}_{j j^\prime j j^\prime} + 2 \mathfrak{C}_{j j^\prime}^{(1)} \mathcal{Z}^{(2)}_{j^\prime j} \}.
\label{eq: C3-2}
\eea

\begin{figure}[th]
\centerline{\includegraphics [width=3.4in, angle=0, clip=on]  {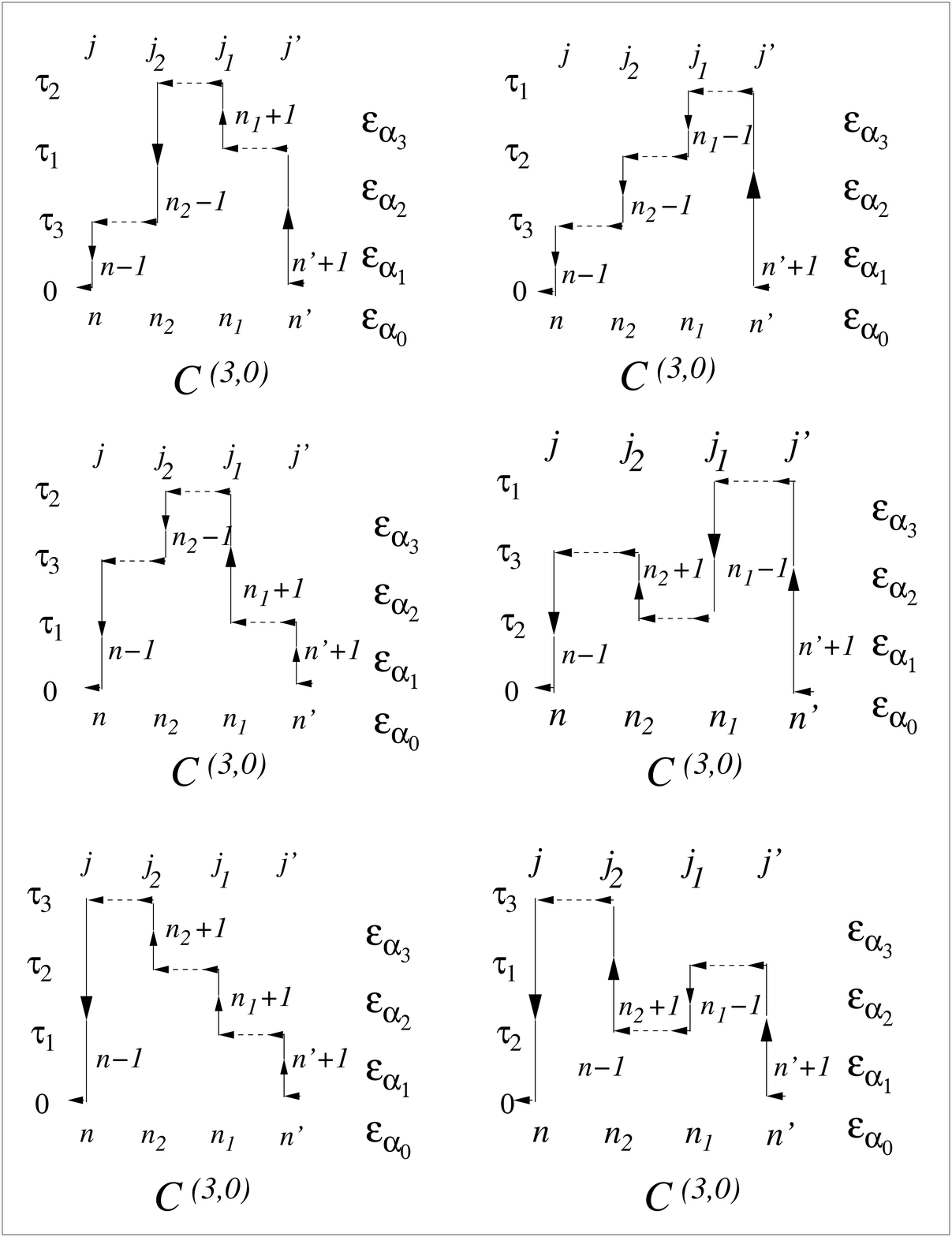}}
\caption[]{(Color online.) Strong-coupling ``diagrams'' for the correlation functions $C^{(3,0)}$. The horizontal directed dashed lines indicate the hopping matrix $\mathfrak{t}$ between the sites labeled, and the vertical lines indicate single-site Green's functions $\mathcal{G}$. The ellipses (red) at multiply visited sites denote the appearance of connected or cumulant Green's functions. Initial and intermediate state labels for the different possible i-time orderings shown are also indicated.}
\label{fig: C-diag-3-0}
\end{figure}

\begin{figure}[th]
\centerline{\includegraphics [width=3.4in, angle=0, clip=on]  {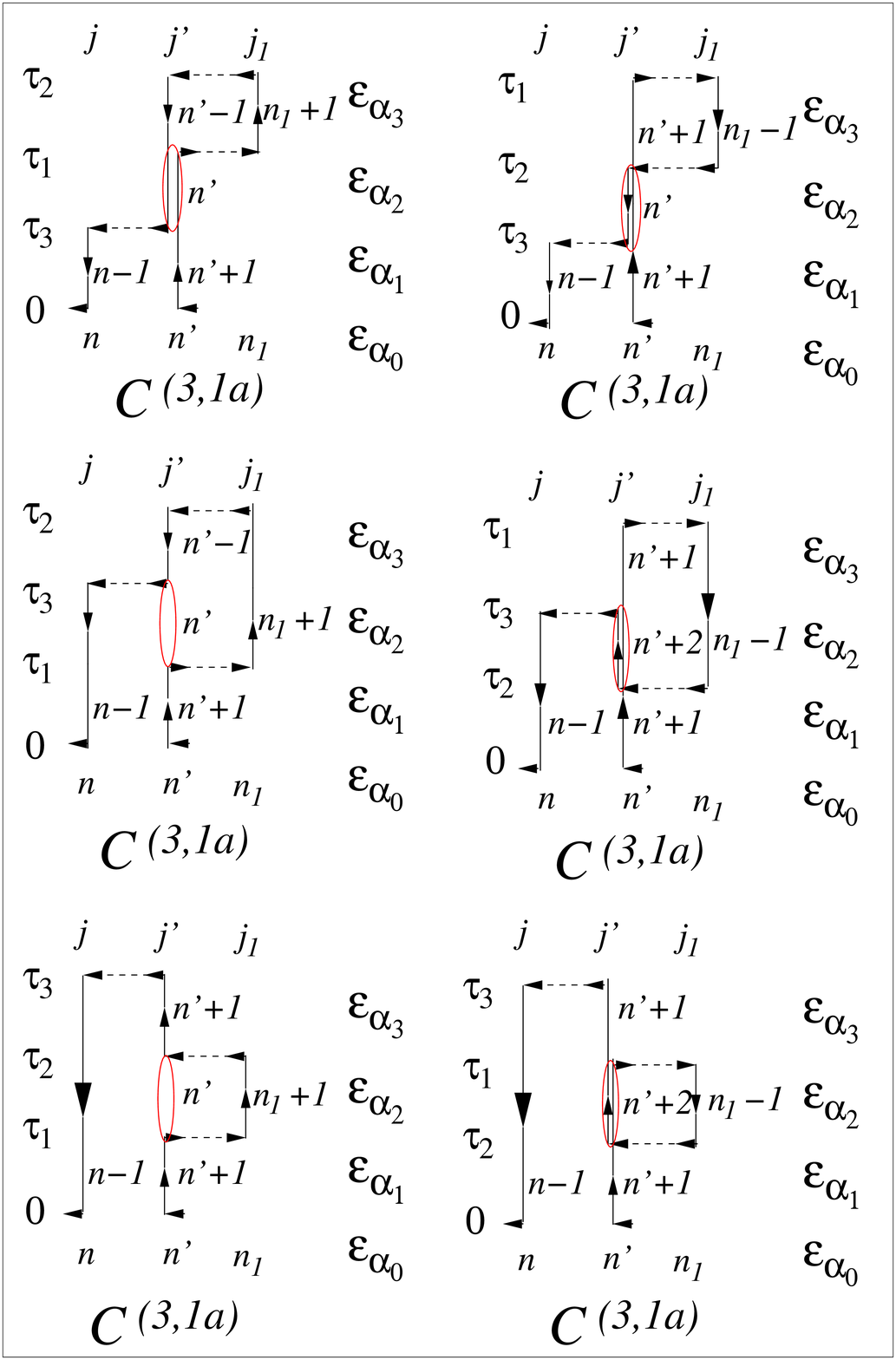}}
\caption[]{(Color online.) Strong-coupling ``diagrams'' for the correlation functions $C^{(3,1)}$. The horizontal directed dashed lines indicate the hopping matrix $\mathfrak{t}$ between the sites labeled, and the vertical lines indicate single-site Green's functions $\mathcal{G}$. The ellipses (red) at multiply visited sites denote the appearance of connected or cumulant Green's functions. Initial and intermediate state labels for the different possible i-time orderings shown are also indicated.}
\label{fig: C-diag-3-1}
\end{figure}

\begin{figure}[th]
\centerline{\includegraphics [width=3.4in, angle=0, clip=on]  {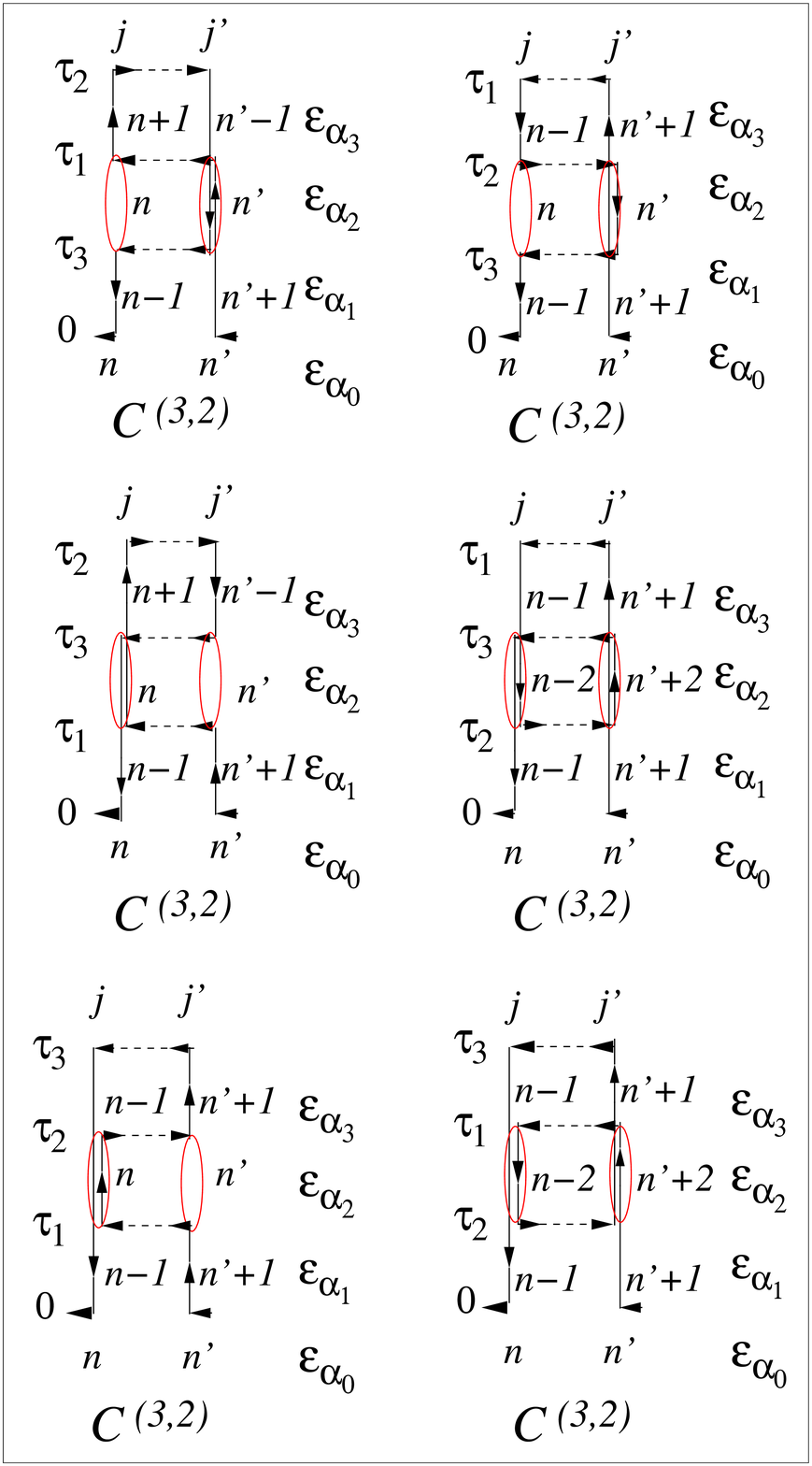}}
\caption[]{(Color online.) Strong-coupling ``diagrams'' for the correlation functions $C^{(3,2)}$. The horizontal directed dashed lines indicate the hopping matrix element $\mathfrak{t}$ between the sites labeled, and the vertical lines indicate single-site Green's functions $\mathcal{G}$. The ellipses (red) at multiply visited sites denote the appearance of connected or cumulant Green's functions. Initial and intermediate state labels for the different possible i-time orderings shown are also indicated.}
\label{fig: C-diag-3-2}
\end{figure}

The i-time integrals that appear in these expressions are most conveniently evaluated by splitting them up into separate integrals corresponding to each of the different ($m!$) i-time orderings of i-time integration variables (in $m^{th}$ order). With each such i-time ordered term one can associate a unique diagram, as shown in Figs.~\ref{fig: C-diag-1}--\ref{fig: C-diag-3-2} up to third order. The diagrams are labeled by the sites that appear, the ``initial'' ($\equiv$ ``final'') and the ``intermediate states'' at these sites as determined by the boson occupation numbers at these sites in each of the i-time intervals, whose labeling corresponds to the boson creation and destruction processes at the sites. (The boson occupation numbers at the sites that do not appear in a diagram do not change with i-time, and play a spectator role, and hence do not appear in the contributions to $C_{j^\prime j}$.)  The ``matrix elements'' that are associated with these processes are then uniquely determined and can be written down by inspection from the labeling. For such a diagram of $m^{th}$ order, let $\mathcal{E}_{\alpha_0}$ and  $\mathcal{E}_{\alpha_1}$, $\cdots$, $\mathcal{E}_{\alpha_m}$ denote the energy eigenvalues of $\mathcal{H}_0$ for the initial (or final) state and the $m$ intermediate states respectively.  Then the i-time integral is of the form,
\bea
&&I_m(\beta; \mathcal{E}_{\alpha_m}, \cdots, \mathcal{E}_{\alpha_1}, \mathcal{E}_{\alpha_0}) = \frac {e^{-\beta\mathcal{E}_{\alpha_0}}} {\mathcal{Z}}  \int_0^{\beta} d\tau_m \int_0^{\tau_m} d\tau_{m-1} \cdots \int_0^{\tau_2} d\tau_{1} \non \\
&\times& \; e^{[\tau_m(\mathcal{E}_{\alpha_0}-\mathcal{E}_{\alpha_m}) + \tau_{m-1} (\mathcal{E}_{\alpha_m}-\mathcal{E}_{\alpha_{m-1}}) + \cdots + \tau_2 (\mathcal{E}_{\alpha_3}-\mathcal{E}_{\alpha_2}) + \tau_1 (\mathcal{E}_{\alpha_2} -\mathcal{E}_{\alpha_1}) ]}
\eea
This is easily evaluated using Laplace-transform techniques, as shown in Appendix~\ref{appendixa}. If the energies are all distinct, then one finds that the integral is the following sum of $m+1$ terms.
\beq
I_m(\beta; \mathcal{E}_{\alpha_m}, \cdots, \mathcal{E}_{\alpha_1}, \mathcal{E}_{\alpha_0}) =  \sum_{\ell = 0}^m \frac {e^{-\beta \mathcal{E}_{\alpha_\ell}}} {\mathcal{Z}} \prod_{\ell^\prime \ne \ell} \frac {1}{(\mathcal{E}_{\alpha_{\ell^\prime}}-\mathcal{E}_{\alpha_\ell})}.
\label{eq: I_m-0}
\eeq
Note that only energy differences appear in the energy denominators in this expression, and they are related in a simple way to the boson creation and destruction processes at the sites that appear in the diagrams; these can be written down by inspection from the labeling shown in each diagram. As the initial and intermediate states at all the sites that do not appear in the diagrams are constrained to be the same, one can replace the Boltzmann factors for the initial and intermediate states by a product of the density matrices {\em for just the sites that appear in the diagrams}.

The expression in Eq.~(\ref{eq: I_m-0}) is non-singular and remains well defined even when one or more of the energies $\mathcal{E}_{\alpha_0}, \mathcal{E}_{\alpha_1}, \cdots, \mathcal{E}_{\alpha_m}$ become equal, as clearly happens, for example, in the diagrams for $C^{(2,1)}$ (see Fig.~\ref{fig: C-diag-1}). For example, if one and only one pair of energies are equal, say, $\mathcal{E}_{\alpha_r} = \mathcal{E}_{\alpha_p}$, then instead of Eq.~(\ref{eq: I_m-0}) one should use the expression
\beq
\sum_{\ell \ne r,p} \frac {e^{-\beta \mathcal{E}_{\alpha_\ell}}} {\mathcal{Z}} \prod_{\ell^\prime \ne \ell} \frac {1}{(\mathcal{E}_{\alpha_{\ell^\prime}}-\mathcal{E}_{\alpha_\ell})} + \frac {e^{-\beta \mathcal{E}_{\alpha_r}}} {\mathcal{Z}} [\beta - \sum_{\ell \ne r,p} \frac{1}{(\mathcal{E}_{\alpha_{\ell}}-\mathcal{E}_{\alpha_r})}] \prod_{\ell^\prime \ne r,p} \frac {1}{(\mathcal{E}_{\alpha_{\ell^\prime}}-\mathcal{E}_{\alpha_r})},
\label{eq: I_m-1}
\eeq
which it reduces to in this limit (see Appendix~\ref{appendixa} for further details).

The diagrams labeled $C^{(2,0)}$ in Fig.~\ref{fig: C-diag-1} shows the two diagrams for $\mathfrak{C}_{j j_1 j^\prime}^{(2)}$, corresponding to the two i-time orderings $ \tau_2 > \tau_1 $ and $ \tau_1 > \tau_2 $ . Each diagram gives rise to three contributions as pointed out above, which are easily written down using the above rules, leading to
\bea
\mathfrak{C}_{j j_1 j^\prime}^{(2)} & = &  \mathfrak{t}_{j j_1} \mathfrak{t}_{j_1 j^\prime} \, \sum_{n, n_1, n^\prime}  \{ \; n (n_1 + 1)(n^\prime + 1) \; [  \frac{\rho_{j, n} \,\rho_{j_1, n_1} \, \rho_{j^\prime, n^\prime}}{(\epsilon^-_{j, n} +  \epsilon^+_{j_1, n_1})(\epsilon^-_{j, n} + \epsilon^+_{j^\prime, n^\prime})} \non \\ & + & \frac{\rho_{j, n-1} \, \rho_{j_1, n_1} \, \rho_{j^\prime, n^\prime + 1}} {(\epsilon^+_{j, n-1} +  \epsilon^-_{j^\prime, n^\prime + 1})(\epsilon^+_{j_1, n_1} +  \epsilon^-_{j^\prime, n^\prime + 1})}  +  \frac{\rho_{j, n-1} \, \rho_{j_1, n_1 + 1} \, \rho_{j^\prime, n^\prime }} {(\epsilon^+_{j, n-1} +  \epsilon^-_{j_1, n_1 + 1})(\epsilon^-_{j_1, n_1 + 1} +  \epsilon^+_{j^\prime, n^\prime })}  ] \non \\
&  & \; \; + \;\; n n_1 (n^\prime + 1) \; [ \frac{\rho_{j, n} \,\rho_{j_1, n_1} \, \rho_{j^\prime, n^\prime}}{(\epsilon^-_{j, n} +  \epsilon^+_{j^\prime, n^\prime})(\epsilon^-_{j_1, n_1} +  \epsilon^+_{j^\prime, n^\prime})} \non \\
& + & \frac{\rho_{j, n-1} \, \rho_{j_1, n_1} \, \rho_{j^\prime, n^\prime + 1}} {(\epsilon^+_{j, n-1} + \epsilon^-_{j_1, n_1})(\epsilon^+_{j, n-1} +  \epsilon^-_{j^\prime, n^\prime + 1})} +  \frac{\rho_{j, n} \, \rho_{j_1, n_1 - 1} \, \rho_{j^\prime, n^\prime + 1 }} {(\epsilon^-_{j, n} +  \epsilon^+_{j_1, n_1 - 1})(\epsilon^+_{j_1, n_1 - 1} +  \epsilon^-_{j^\prime, n^\prime + 1 })}  ] \; \}
\label{eq: C2-jj1jp-1}\\
& = & \mathfrak{t}_{j j_1} \mathfrak{t}_{j_1 j^\prime} \, \sum_{n, n_1, n^\prime}  \; \rho_{j, n} \,\rho_{j_1, n_1} \, \rho_{j^\prime, n^\prime}  \; \{ \; [  \frac{ n (n_1 + 1)(n^\prime + 1)}{(\epsilon^-_{j, n} +  \epsilon^+_{j_1, n_1})(\epsilon^-_{j, n} + \epsilon^+_{j^\prime, n_j^\prime})} \non \\ & + & \frac{ (n + 1) (n_1 + 1) n^\prime } {(\epsilon^+_{j, n} +  \epsilon^-_{j^\prime, n^\prime}) (\epsilon^+_{j_1, n_1} +  \epsilon^-_{j^\prime, n^\prime})}  +  \frac{ (n + 1) n_1 (n^\prime + 1)} {(\epsilon^+_{j, n} +  \epsilon^-_{j_1, n_1})(\epsilon^-_{j_1, n_1} +  \epsilon^+_{j^\prime, n^\prime })}  ] \non \\
&  & \; \; + \;\;  \; [ \frac{n n_1 (n^\prime + 1)}{(\epsilon^-_{j, n} +  \epsilon^+_{j^\prime, n^\prime})(\epsilon^-_{j_1, n_1} +  \epsilon^+_{j^\prime, n^\prime})} \non \\
& + & \frac{(n + 1) n_1 n^\prime } {(\epsilon^+_{j, n} + \epsilon^-_{j_1, n_1})(\epsilon^+_{j, n} +  \epsilon^-_{j^\prime, n^\prime})} +  \frac{n (n_1 + 1) n^\prime } {(\epsilon^-_{j, n} +  \epsilon^+_{j_1, n_1 })(\epsilon^+_{j_1, n_1 } +  \epsilon^-_{j^\prime, n^\prime })}  ]  \; \}.
\label{eq: C2-jj1jp-2}
\eea
Again, the second form of the result, Eq.~(\ref{eq: C2-jj1jp-2}), is obtained by appropriately relabeling the bosonic occupation  numbers in four of the six terms in the first form [Eq.~(\ref{eq: C2-jj1jp-1})], and is easier to use at $T=0$.

Similarly, from the diagrams contributing to $\bar{\mathfrak{C}}_{j j_1 j}^{(2)}$ (labeled $C^{(2,1)}$) and $\mathcal{Z}^{(2)}_{j j_1}$ (labeled $Z^{(2)}$) shown in Fig.~\ref{fig: C-diag-1}, and using Eq.~(\ref{eq: I_m-1}), we obtain, for the two equivalent forms for each,
\bea
\bar{\mathfrak{C}}_{j j_1 j}^{(2)} & = &    \mathfrak{t}_{j j_1} \mathfrak{t}_{j_1 j} \, \sum_{n, n_1}  (n) \times \non \\
& \{ &  n (n_1 + 1) \; [ \frac {\rho_{j, n} \, \rho_{j_1, n_1}} {(\epsilon^-_{j, n} +  \epsilon^+_{j_1, n_1})} ( \beta -  \frac {1}  {(\epsilon^-_{j, n} +  \epsilon^+_{j_1, n_1})} ) + \frac{\rho_{j, n-1} \,\rho_{j_1, n_1 + 1} }{(\epsilon^+_{j, n-1} +  \epsilon^-_{j_1, n_1+1})^2} ] \non \\
& + &  (n + 1) n_1  \; [ \frac { \rho_{j, n} \, \rho_{j_1, n_1}} {(\epsilon^+_{j, n} +  \epsilon^-_{j_1, n_1})} ( \beta - \frac {1}  {(\epsilon^+_{j, n} +  \epsilon^-_{j_1, n_1})} ) + \frac {\rho_{j, n + 1} \,\rho_{j_1, n_1 - 1}} {(\epsilon^-_{j, n+1} + \epsilon^+_{j_1, n_1 - 1})^2} ] \; \}
\label{eq: C2-jj1j-1} \\
& = &   \mathfrak{t}_{j j_1} \mathfrak{t}_{j_1 j} \, \sum_{n, n_1}   \rho_{j, n} \, \rho_{j_1, n_1}   \times \non \\
& \{ &  [  \frac {n (n_1 + 1) n}{(\epsilon^-_{j, n} +  \epsilon^+_{j_1, n_1})} ( \beta -  \frac {1}  {(\epsilon^-_{j, n} +  \epsilon^+_{j_1, n_1})} ) + \frac{(n + 1 ) n_1 (n + 1)}{(\epsilon^+_{j, n} +  \epsilon^-_{j_1, n_1})^2}  ] \non \\
& + &  [ \frac {(n + 1) n_1 n} {(\epsilon^+_{j, n} +  \epsilon^-_{j_1, n_1})} ( \beta   -  \frac {1}  {(\epsilon^+_{j, n} +  \epsilon^-_{j_1, n_1})} ) + \frac {n (n_1 + 1) (n - 1)} {(\epsilon^-_{j, n} +  \epsilon^+_{j_1, n_1})^2} ] \; \}
\label{eq: C2-jj1j-2}
\eea
\bea
\mathcal{Z}^{(2)}_{j j_1} & = &   \mathfrak{t}_{j j_1} \mathfrak{t}_{j_1 j} \, \sum_{n, n_1} \{ \;  n (n_1 + 1) \; [ \frac {\rho_{j, n} \, \rho_{j_1, n_1}} {(\epsilon^-_{j, n} +  \epsilon^+_{j_1, n_1})} ( \beta -  \frac {1}  {(\epsilon^-_{j, n} +  \epsilon^+_{j_1, n_1})} ) + \frac{\rho_{j, n-1} \,\rho_{j_1, n_1 + 1} }{(\epsilon^+_{j, n-1} +  \epsilon^-_{j_1, n_1+1})^2} ] \non \\
& + &  (n + 1) n_1  \; [ \frac { \rho_{j, n} \, \rho_{j_1, n_1}} {(\epsilon^+_{j, n} +  \epsilon^-_{j_1, n_1})} ( \beta - \frac {1}  {(\epsilon^+_{j, n} +  \epsilon^-_{j_1, n_1})} ) + \frac {\rho_{j, n + 1} \,\rho_{j_1, n_1 - 1}} {(\epsilon^-_{j, n+1} +  \epsilon^+_{j_1, n_1 - 1})^2} ] \; \}
\label{eq: Z2-jj1-1}\\
& =  &   \mathfrak{t}_{j j_1} \mathfrak{t}_{j_1 j} \, \sum_{n, n_1}   \rho_{j, n} \, \rho_{j_1, n_1}   \times \non \\
& \{ &  [\frac { n (n_1 + 1)}{(\epsilon^-_{j, n} +  \epsilon^+_{j_1, n_1})} ( \beta -  \frac {1}  {(\epsilon^-_{j, n} +  \epsilon^+_{j_1, n_1})} ) + \frac{(n + 1) n_1 }{(\epsilon^+_{j, n} +  \epsilon^-_{j_1, n_1})^2} ] \non \\
& + &[  \frac {(n + 1) n_1} {(\epsilon^+_{j, n} +  \epsilon^-_{j_1, n_1})} ( \beta   -  \frac {1}  {(\epsilon^+_{j, n} +  \epsilon^-_{j_1, n_1})} ) + \frac { n (n_1 + 1)} {(\epsilon^-_{j, n} +  \epsilon^+_{j_1, n_1})^2} ] \; \}.
\label{eq: Z2-jj1-2}
\eea
Simplifying these expressions we find
\bea
\bar{\mathfrak{C}}_{j j_1 j}^{(2)}  =  \mathfrak{t}_{j j_1} \mathfrak{t}_{j_1 j} \, \sum_{n, n_1}   \rho_{j, n} \, \rho_{j_1, n_1} & \{ & \beta \; n \; [ \frac {n (n_1 + 1)}{(\epsilon^-_{j, n} +  \epsilon^+_{j_1, n_1})} +  \frac {(n + 1) n_1} {(\epsilon^+_{j, n} +  \epsilon^-_{j_1, n_1})} ] \non \\
& + &  [ \frac{1}{(\epsilon^+_{j, n} +  \epsilon^-_{j_1, n_1})^2} - \frac {1} {(\epsilon^-_{j, n} +  \epsilon^+_{j_1, n_1})^2} ] \; \}
\label{C2-jj1j}
\eea
\beq
\mathcal{Z}^{(2)}_{j j_1} = \mathfrak{t}_{j j_1} \mathfrak{t}_{j_1 j} \, \sum_{n, n_1}   \rho_{j, n} \, \rho_{j_1, n_1} \{ \; \beta \; [ \frac {n (n_1 + 1)}{(\epsilon^-_{j, n} +  \epsilon^+_{j_1, n_1})} +  \frac {(n + 1) n_1} {(\epsilon^+_{j, n} +  \epsilon^-_{j_1, n_1})} ] \; \}.
\label{eq: Z2-jj1}
\eeq

Finally, Figs.~\ref{fig: C-diag-3-0}, \ref{fig: C-diag-3-1} and \ref{fig: C-diag-3-2} show the diagrams corresponding respectively to $\mathfrak{C}^{(3)}_{j j_2 j_1 j^\prime}$, $\bar{\mathfrak{C}}^{(3)}_{j j^\prime j_1 j^\prime}$ (the digrams for $\bar{\mathfrak{C}}^{(3)}_{j j_2 j j^\prime}$ can be obtained from those in Fig.~\ref{fig: C-diag-3-1} by symmetry and a simple relabeling) and $\bar{\mathfrak{C}}^{(3)}_{j j^\prime j j^\prime}$. (For simplicity, since we do not discuss nonbipartite lattices in detail in this paper, the diagrams for $\bar{\mathfrak{C}}^{(3)}_{j j_2 j_1 j} $ are not shown.) In each case, there are 6 possible orderings of the i-time variables $\tau_3, \tau_2$ and $\tau_1$ [listed in the order $(\tau_3 > \tau_2 > \tau_1)$, $(\tau_2 > \tau_3 > \tau_1)$, $(\tau_2 > \tau_1 > \tau_2)$, $(\tau_3 > \tau_1 > \tau_2)$, $(\tau_1 > \tau_3 > \tau_2)$, and $(\tau_1 > \tau_2 > \tau_3)$ below]; and from each i-time ordering we get four contributions corresponding to the initial and three intermediate states (apart from the subtractions arising from the connected two-particle Green's functions). The contributions can be written down straightforwardly using the rules stated above, and we present them in Appendix~\ref{appendixb}.

It is easy to see that the methods we have discussed above permit one, in principle, to similarly write down the contributions to $G_{j j^\prime}$ and $C_{j^\prime j}$ to higher orders as well, though the calculations will become increasingly tedious {\em unless one can find a way to automate them}. However, it is possible to calculate sums of subsets of these contributions to arbitrary orders.

The easiest subset to sum is $G_{j j^\prime}^{(m,0)}$, with contributions coming entirely from ``self-avoiding lattice walks'' while computing thermal averages, and therefore involving only single-particle single-site Green's functions, {\em but ignoring the self-avoidance constraint while summing over the different possible walks}. From the above analysis it is clear that the resulting term is
\bea
G_{jj^\prime}^{(m;0)}(\tau, \tau^\prime) & = & (-1)^{(m+1)}  \mathfrak{t}_{j j_{m-1}} \mathfrak{t}_{j_{m-1} j_{m-1}} \cdots \mathfrak{t}_{j_2 j_1} \mathfrak{t}_{j_1 j^\prime} \int_{\tau_m} \int_{\tau_{m-1}} \cdots \int_{\tau_2} \int_{\tau_1} \non \\
& \times & \mathcal{G}_{j}(\tau,\tau_m)  \mathcal{G}_{j_{m-1}}(\tau_m,\tau_{m-1}) \cdots \mathcal{G}_{j_1}(\tau_2,\tau_1) \mathcal{G}_{j^\prime}(\tau_1, \tau^\prime).
\label{eq: Gm-0}
\eea
The sum of these to arbitrary order, together with $G_{j j^\prime}^{(1)}$ [Eq.~(\ref{eq: G1})] and $G_{j j^\prime}^{(0)}$ [Eq.~(\ref{eq: G0})], correspond to a geometric series for the Green's function regarded as a matrix (denoted by bold-face letters and/or square brackets below) with lattice sites and i-times as indices, and corresponds to the well known ``Random Phase Approximation'' (RPA) result,
\beq
[\mathbf{G}^{RPA}]^{-1}_{j j^\prime}(\tau, \tau^\prime) =   [\mathbf {\mathcal{G}}_j]^{-1}(\tau, \tau^\prime) \delta_{j j^\prime} +  \delta (\tau, \tau^\prime) \mathfrak{t}_{j j^\prime}.
\label{eq: G-RPA}
\eeq
By taking a Fourier transform with respect to the even Matsubara frequencies $i\Omega_m \equiv 2 m \pi k_B T  $, $ m = 0, \pm 1, \pm 2, \cdots$, one can write this as a matrix equation involving only lattice indices:
\beq
[\mathbf{G}^{RPA}]^{-1}_{j j^\prime}(i\Omega_m) =   [\mathcal{G}_j(i\Omega_m)]^{-1} \delta_{j j^\prime} + \mathfrak{t}_{j j^\prime},
\label{eq: G-omega-RPA}
\eeq
where, from Eq.~(\ref{eq: calgI}), the single-site Green's function at a fixed frequency is easily obtained as
\beq
\mathcal{G}_j(i\Omega_m) = \sum_{n} \rho_{j, n} \; [ \frac {(n +1)} {(i\Omega_m - \epsilon^+_{j, n})} \, - \frac  {n} { (i\Omega_m + \epsilon^-_{j, n})} ].
\label{eq: calgI-w}
\eeq
The RPA correlation function can then be straightforwardly obtained, in view of Eq.~(\ref{eq: CfromG}), as the Matsubara frequency sum
\beq
C_{j^\prime j} = - [\mathbf{G}]_{j j^\prime}(0, 0^+) = -\sum_{\Omega_m} G_{j j^\prime}(i\Omega_m) e^{i\Omega_m 0^+} \, ,
\label{eq: CfromG-freq}
\eeq
by using $[\mathbf{G}^{RPA}]$ for $[\mathbf{G}]$. While the frequency sum can in principle be evaluated using standard contour integral techniques~\cite{mahan}, in the inhomogeneous case the above calculation involves a matrix inversion with respect to the site indices.

The calculations simplify, however, for the case of a homogeneous system, {\it i.~e.}, without a trap or disorder potential, and for $T \rightarrow 0$. Then all the sites are identical, and in a generic case, the ground state of $\mathcal{H}_{0j}$ at every site corresponds to the same fixed boson occupancy which we denote $n_{}$. In this limit, one has $\epsilon_{n}  = -\mu n + \frac{U}{2} {n}({n}-1)$ whence $\epsilon^+_{n_{}} = -\mu + U n_{}$ and  $\epsilon^-_{n_{}} = \mu - U (n_{} - 1) $. The RPA momentum distribution can be calculated exactly analytically in this case, as discussed in detail by Sengupta and Dupuis~\cite{rpa}. From Eq.~(\ref{eq: calgI-w}), at $T \ll \epsilon^\pm_{n_{}}$ we get, for all sites $j$,
\beq
\mathcal{G}_j(i\Omega_m) =  \left [ \frac {(n_{} +1)} {(i\Omega_m - \epsilon^+_{n_{}})} \, - \frac  {n_{}} { (i\Omega_m + \epsilon^-_{n_{}})} \right ].
\label{eq: calgI-kw-0T}
\eeq
Hence,
\bea
[\mathbf{G}^{RPA}]_{\bf{k}}(i\Omega_m) & = & \frac {1} {[\mathcal{G}_j(i\Omega_m)]^{-1} - \epsilon_{\bf{k}}} \non \\
& = & \frac {1-z_{\bf{k}}}{(i\Omega_m -E^-_{\bf{k}})} + \frac {z_{\bf{k}}}{(i\Omega_m -E^+_{\bf{k}})}
\label{eq: G-omega-k-RPA}
\eea
with poles at $ E^\pm_{\bf{k}} \equiv [\epsilon_{\bf{k}}+ \epsilon^+_{n_{}} - \epsilon^-_{n_{}} \pm \sqrt{\epsilon_{\bf{k}}^2 + 2 \epsilon_{\bf{k}} U (2n_{} + 1) + U^2}] / 2 $ and residues determined in terms of $z_{\bf{k}} \equiv  (E^+_{\bf{k}} + \mu + U) / (E^+_{\bf{k}}- E^-_{\bf{k}})$.  The RPA momentum distribution at $T=0$ is just the negative of the spectral weight of the pole at $E^-_{\bf{k}}$; {\it i.~e.},
\beq
n_{\bf{k}}^{RPA} = z_{\bf{k}} - 1 = \frac {E^-_{\bf{k}} + \mu + U} {\sqrt{\epsilon_{\bf{k}}^2 +2 \epsilon_{\bf{k}}U(2n_{}+1)+ U^2}}.
\eeq

The challenge, of course, is to go beyond the RPA. One way to achieve this, by summing further infinite subsets of
contributions to $G_{j j^\prime}$ beyond the RPA, is by using the Dyson equation (compare Eq.~\ref{eq: G-RPA}),
\beq
[\mathbf{G}]^{-1}_{j j^\prime}(\tau, \tau^\prime) =   [\mathbf{\mathcal{G}}_j]^{-1}(\tau, \tau^\prime) \delta_{j j^\prime} +  \delta (\tau, \tau^\prime) \mathfrak{t}_{j j^\prime} - \mathbf{\Sigma}^{(2)}_{j j^\prime}(\tau, \tau^\prime) - \mathbf{\Sigma}^{(3)}_{j j^\prime}(\tau, \tau^\prime) - \cdots \, ,
\label{eq: G-Dyson}
\eeq
where $\mathbf{\Sigma}^{(m)}$ denotes a self-energy correction that corrects RPA to order $\mathfrak{t}^m$. By re-expanding the inverse of this equation and comparing with the expansion for $G_{j j^\prime}$ discussed earlier, it is straightforward to obtain the following expressions for the self-energy corrections up to third order in $\mathfrak{t}$.
\bea
\mathbf{\Sigma}^{(2)}_{j j^\prime}(\tau, \tau^\prime) & = & \int_{\tau_2} \int_{\tau_1} [\mathbf {\mathcal{G}}_j]^{-1}(\tau, \tau_2) G_{jj^\prime}^{(2;1)}(\tau_2, \tau_1) [\mathbf {\mathcal{G}}_{j^\prime}]^{-1}(\tau_1, \tau^\prime) \non \\
& = & \delta_{j j^\prime} \int_{\tau_2} \int_{\tau_1} \sum_{j_1} [\mathbf {\mathcal{G}}_j]^{-1}(\tau, \tau_2)\tilde{\mathfrak{G}}^{(2)}_{j j_1 j}(\tau_2, \tau_1) [\mathbf {\mathcal{G}}_j]^{-1}(\tau_1, \tau^\prime),
\label{eq: sig-2}
\eea
with $\tilde{\mathfrak{G}}^{(2)}$ as given in Eq.~(\ref{eq: G2-1}).
Similarly,
\bea
\mathbf{\Sigma}^{(3)}_{j j^\prime}(\tau, \tau^\prime) & = & \int_{\tau_2} \int_{\tau_1}  [\mathbf {\mathcal{G}}_j]^{-1}(\tau, \tau_2) [ G_{jj^\prime}^{(3;1)}(\tau_2, \tau_1) + G_{jj^\prime}^{(3;2)}(\tau_2, \tau_1) + G_{jj^\prime}^{(3;3)}(\tau_2, \tau_1) ] [\mathbf {\mathcal{G}}_{j^\prime}]^{-1}(\tau_1, \tau^\prime)  \non \\
& + & \int_{\tau_1} \sum_{j_2} \mathfrak{t}_{j j_2} G_{j_2 j^\prime}^{(2;1)}(\tau, \tau_1) [\mathbf {\mathcal{G}}_{j^\prime}]^{-1}(\tau_1, \tau^\prime)  + \int_{\tau_2} \sum_{j_2} [\mathbf {\mathcal{G}}_j]^{-1}(\tau, \tau_2) G_{j j_2}^{(2;1)}(\tau_2, \tau^\prime) \mathfrak{t}_{j_2 j^\prime}. \non
\eea
It is straightforward to verify using the expressions given in Eqs.~(\ref{eq: G2-1}) and (\ref{eq: G3-1})--(\ref{eq: G3-3}), that the term involving $G_{jj^\prime}^{(3;1)}$ exactly cancels the two terms involving $G_{jj^\prime}^{(2;1)}$ above, and one obtains,
\beq
\mathbf{\Sigma}^{(3)}_{j j^\prime}(\tau, \tau^\prime) =  - \int_{\tau_2} \int_{\tau_1}  [\mathbf {\mathcal{G}}_j]^{-1}(\tau, \tau_2) [ \frac{1}{2!} \tilde{\mathfrak{G}}^{(3)}_{j j^\prime j j^\prime}(\tau_2, \tau_1) + \delta_{j j^\prime} \sum_{j_2, j_1} \tilde{\mathfrak{G}}^{(3)}_{j j_2 j_1 j} (\tau_2, \tau_1)] [\mathbf {\mathcal{G}}_{j^\prime}]^{-1}(\tau_1, \tau^\prime).
\label{eq: sig-3-final}
\eeq
One can in principle evaluate these expressions for the self-energies explicitly using the techniques discussed above, and thereby determine spectral functions as well as the momentum distribution function using Eq.~(\ref{eq: CfromG-freq}). We plan to complete such work in the future.

However, in this paper we adopt a different procedure for calculating the momentum distribution function for the homogeneous case and in the $T \rightarrow 0$ limit.  We directly evaluate the expressions for $C_{j^\prime j}$ up to third order in $\mathfrak{t}$. Then we use a scaling ansatz for the momentum distribution function determined in such a way that when expanded in powers of $\mathfrak{t}$, it agrees with our calculated results, thereby effecting an infinite order resummation in a different way which automatically has the correct critical behavior at the Mott-superfluid transition.

The direct evaluation of our expressions for $C_{j^\prime j}$ in the homogeneous, $T \rightarrow 0$ limit is straightforward. The $T \rightarrow 0$ limit is easiest to implement using {\em the second form of these expressions}, where, just as discussed above in case of the RPA,  for $T \ll \epsilon^\pm_{n_{}}$ the sums over the initial and intermediate states are all restricted to $n_{}$. The excitation energies that occur in the energy denominators in these expressions are given by $(\epsilon^+_{n_{}} +  \epsilon^-_{n_{}}) = U $, $\epsilon^+_{n_{}} + \epsilon^-_{n_{} - 1} + \epsilon^+_{n_{}} +  \epsilon^-_{n_{}} = \epsilon^+_{n_{}} + \epsilon^-_{n_{}} + \epsilon^+_{n_{} + 1} +  \epsilon^-_{n_{}} = 3 U $ and $\epsilon^+_{n_{}} + \epsilon^-_{n_{} - 1} + \epsilon^+_{n_{} + 1} +  \epsilon^-_{n_{}} = 4 U $. Hence, we find
\bea
C^{(0)}_{j^\prime j} & = & \delta_{j, j^\prime} n_{} \\
C^{(1)}_{j^\prime j} & = &  \mathfrak{C}^{(1)}_{j j^\prime} =\mathfrak{t}_{j j^\prime} \left  \{ \frac{2n_{} (n_{} + 1)}{U} \right \}.
\eea
For the second-order terms we obtain
\bea
\mathfrak{C}^{(2)}_{j j_1 j^\prime} & = & \mathfrak{t}_{j j_1} \mathfrak{t}_{j_1 j^\prime} \left \{ \left [ \frac{3 n_{} (n_{} + 1)^2}{U^2} \right ] + \left [ \frac{3 n_{}^2 (n_{} + 1)}{U^2} \right ] \right \} \non \\
& = & \mathfrak{t}_{j j_1} \mathfrak{t}_{j_1 j^\prime} \left \{  \frac{3 n_{} (n_{} + 1)(2n_{} + 1)}{U^2} \right \} \\
\bar {\mathfrak{C}}^{(2)}_{j j_1 j} & = & n_{} \; \mathcal{Z}^{(2)}_{j j_1} = n_{} \; \beta \; \mathfrak{t}_{j j_1} \mathfrak{t}_{j_1 j} \left \{ \frac{2 n_{} (n_{} + 1)}{U}  \right \}.
\eea
Hence, using Eqs.~(\ref{eq: C2-0-1}) and (\ref{eq: C2-1-1}), we find
\bea
C^{(2,0)}_{j^\prime j} & = & - \left [ \sum_{j_1} \mathfrak{t}_{j j_1} \mathfrak{t}_{j_1 j^\prime}\right ] \left \{  \frac{3 n_{} (n_{} + 1)(2n_{} + 1)}{U^2} \right \} \\
C^{(2,1)}_{j^\prime j} & = & - \delta_{j, j^\prime} C^{(2,0)}_{j j}.
\eea
Note that in $C^{(2,1)}$ the (divergent) temperature dependent terms from $\bar {\mathfrak{C}}^{(2)}$ and $\mathcal{Z}^{(2)}$ exactly cancel, as they ought to.

Finally, we consider the various third-order terms. We get the following results
\bea
\mathfrak{C}^{(3)}_{j j_2 j_1 j^\prime} & = & \mathfrak{t}_{j j_2} \mathfrak{t}_{j_2 j_1} \mathfrak{t}_{j_1 j^\prime} \left \{ \left [ \frac{4 n_{} (n_{} + 1)^3}{U^3}\right ] +  4 \left [\frac{3 n_{}^2 (n_{} + 1)^2}{U^3}\right ] + \left [ \frac{4 n_{}^3 (n_{} + 1)}{U^3}\right ] \right \} \\
& = &  \mathfrak{t}_{j j_2} \mathfrak{t}_{j_2 j_1} \mathfrak{t}_{j_1 j^\prime}  \left \{  \frac{4 n_{} (n_{} + 1)(5 n_{}^2 + 5 n_{} + 1)}{U^3} \right \},
\eea
whence,
\beq
C^{(3,0)}_{j^\prime j}  =  \left [ \sum_{j_2, j_1} \mathfrak{t}_{j j_2} \mathfrak{t}_{j_2 j_1} \mathfrak{t}_{j_1 j^\prime} \right ] \left \{  \frac{4 n_{} (n_{} + 1)(5 n_{}^2 + 5 n_{} + 1)}{U^3} \right \}.
\eeq
Furthermore,
\bea
\bar {\mathfrak{C}}^{(3)}_{j j^\prime j_1 j^\prime} & = & \mathfrak{t}_{j j^\prime} \mathfrak{t}_{j^\prime j_1} \mathfrak{t}_{j_1 j^\prime} \left \{ \left [\frac{2 n_{} (n_{} + 1)^3}{U^3} + \frac{n_{}^2 (n_{} + 1)^2}{U^3} (\beta U - 2)\right ] \right .\non \\
& + &  2 \left [\frac{2 n_{}^2 (n_{} + 1)^2}{U^3} + \frac{n_{}(n_{} + 1)(n_{}^2 - 1)}{3U^3} + \frac{n_{}(n_{} + 1)(n_{}^2 + 2 n_{})}{3U^3}\right ] \non \\
& + & 2 \left [\frac{n_{}^2 (n_{} + 1)^2}{U^3} (\beta U - 2)+ \frac{n_{}(n_{} + 1)(n_{}^2 - 1)}{3U^3} + \frac{n_{}(n_{} + 1)(n_{}^2 + 2 n_{})}{3U^3}\right ] \non \\
& + & \left . \left [\frac{2 n_{}^3 (n_{} + 1)}{U^3} + \frac{n_{}^2 (n_{} + 1)^2}{U^3} (\beta U - 2)\right ] \right \} \\
& = & \mathfrak{t}_{j j^\prime} \mathfrak{t}_{j^\prime j_1} \mathfrak{t}_{j_1 j^\prime} \left \{ \frac{2 n_{} (n_{} + 1)(4 n_{}^2 + 4 n_{} + 1)}{3U^3} + 4 \frac{n_{}^2 (n_{} + 1)^2}{U^2} \; \beta  \right \}.
\eea
Similarly, we find
\beq
\bar {\mathfrak{C}}^{(3)}_{j j_2 j j^\prime}  = \mathfrak{t}_{j j_2} \mathfrak{t}_{j_2 j} \mathfrak{t}_{j j^\prime} \left \{ \frac{2 n_{} (n_{} + 1)(4 n_{}^2 + 4 n_{} + 1)}{3U^3} + 4 \frac{n_{}^2 (n_{} + 1)^2}{U^2} \; \beta  \right \}.
\eeq
Using these and Eq.~(\ref{eq: C3-1}), we get
\beq
C^{(3,1)}_{j^\prime j}  = -  \left [ \sum_{j_1} \mathfrak{t}_{j j^\prime} \mathfrak{t}_{j^\prime j_1} \mathfrak{t}_{j_1 j^\prime} + \sum_{j_2} \mathfrak{t}_{j j_2} \mathfrak{t}_{j_2 j} \mathfrak{t}_{j j^\prime} \right ] \left \{  \frac{2 n_{} (n_{} + 1)( 26 n_{}^2 + 26 n_{} + 5)} {3U^3} \right \}.
\eeq
Next,
\bea
\bar{\mathfrak{C}}^{(3)}_{j j^\prime j j^\prime} & = & \mathfrak{t}_{j j^\prime} \mathfrak{t}_{j^\prime j} \mathfrak{t}_{j j^\prime}  \left \{ 2\left [ 2 \frac{n_{}^2 (n_{} + 1)^2}{U^3} (\beta U - 2) \right ] \right .\non \\
& + &  \left . 4 \left [ \frac{n_{}^2 (n_{} + 1)^2}{U^3} (\beta U - 2) + \frac{4 (n_{} - 1) n_{} (n_{} + 1)(n_{} + 2)}{4U^3} \right ] \right \} \\
& = & \mathfrak{t}_{j j^\prime} \mathfrak{t}_{j^\prime j} \mathfrak{t}_{j j^\prime}  \left \{ 8 \frac{n_{}^2 (n_{} + 1)^2}{U^2} \; \beta -  \frac{2 n_{} (n_{} + 1)(7 n_{}^2 + 7 n_{} + 2)}{U^3} \right \}.
\eea
Hence, using Eq.~(\ref{eq: C3-2}), we obtain
\beq
C^{(3,2)}_{j^\prime j} =  \left [ \mathfrak{t}_{j j^\prime} \mathfrak{t}_{j^\prime j} \mathfrak{t}_{j j^\prime} \right ] \left \{ \frac{ n_{} (n_{} + 1)( 23 n_{}^2 + 23 n_{} + 2)} {3U^3}  \right \}.
\eeq
Note, again, the exact cancelation of the divergent temperature-dependent terms above.
It is straightforward to verify that the Fourier transforms of the expressions for $C^{(m,0)}$ above up to third order agree with those obtainable by expanding the RPA expression in powers of $\epsilon_{\bf{k}}$ (see below).

\section{Scaling analysis}

In the rest of this manuscript, we specialize to the case of nearest-neighbor hopping on a hypercubic lattice in $d$-dimensions. Combining the different contributions for $C_{j^\prime j}$ for such a lattice and Fourier transforming to momentum space, we arrive at the starting point for the scaling analysis, which is the strong-coupling expansion for the zero-temperature momentum distribution truncated to third order in the hopping and shown in Eq.~(\ref{eq: third_order_general}).
It is more convenient to reexpress the results for different cases in terms of the dimensionless parameters $x=d\mathfrak{t}/U$ and
$\xi_{\bf k}=\epsilon_{\bf k}/2d\mathfrak{t}$.  If we further consider only the $n_{}=1$ Mott insulator, we find
\beq
n_{\bf k}=1 - 8 \xi_{\bf k} x + \left [ 72 \xi_{\bf k}^2 - \frac {36}{d} \right ] x^2  - 32 \left [ 22\xi_{\bf k}^3 - \frac{19}{d} + \frac{2}{d^2} \right ] x^3 ;
\eeq
{\it i.~e.},
\begin{equation}
 n_{\bf k}=1-8\xi_{\bf k}x+72\xi_{\bf k}^2x^2-704\xi_{\bf k}^3x^3,
\label{eq: infd}
\end{equation}
in infinite dimensions where $x$ remains finite as $d\rightarrow\infty$,
\begin{equation}
 n_{\bf k}=1-8\xi_{\bf k}x+12[6\xi_{\bf k}^2-1]x^2-32[22\xi_{\bf k}^2-\frac{55}{9}]\xi_{\bf k}x^3,
\label{eq: 3d}
\end{equation}
in three dimensions,
\begin{equation}
 n_{\bf k}=1-8\xi_{\bf k}x+18[4\xi_{\bf k}^2-1]x^2-32[22\xi_{\bf k}^2-9]\xi_{\bf k}x^3,
\label{eq: 2d}
\end{equation}
in two dimensions,
and
\begin{equation}
 n_{\bf k}=1-8\xi_{\bf k}x+36[2\xi_{\bf k}^2-1]x^2-32[22\xi_{\bf k}^2-17]\xi_{\bf k}x^3,
\label{eq: 1d}
\end{equation}
in one dimension.  Note that because integrals of odd powers of $\xi_{\bf k}$ over momentum vanish, we can use the fact that the integral of the square of $\xi_{\bf k}$ over ${\bf k}$ is equal to $1/(2d)$ to show that the integral of the
strong-coupling expansion for $n_{\bf k}$ over ${\bf k}$ is always equal to 1, as it must be.

We show comparison of these truncated third-order strong-coupling expansions directly with exact numerical results and other analytic approximations below.  It turns out that the truncated strong-coupling expansion does not work so well for the momentum distribution once the hopping is on the order of one fourth of the critical hopping for the Mott to superfluid transition in two and three dimensions (and is even worse in one dimension).  Hence, we use additional knowledge about the momentum distribution and how it scales near the critical point, along with the exact solution in large dimensions to create a phenomenological ansatz for the momentum distribution which produces analytical expressions useful for direct comparison with experiment.

We start our scaling analysis with a general discussion.  The zero momentum distribution function becomes critical
at the critical value of the hopping for the Mott insulator to superfluid transition (called $x_c$).  The critical behavior goes like
$n_{\bf k=0}\rightarrow \xi^{(1-\eta)}$ where $\xi\propto 1/(x_c-x)^\nu$ is the correlation length of a $d+1$ dimensional XY model~\cite{fisher_etal} and $\eta$ and $\nu$ are critical exponents in the usual notation~\cite{chaikin}.   In two and higher dimensions, the correlation length
diverges as a power law.  The critical exponents for the two-dimensional Bose Hubbard model, which correspond to the
three-dimensional XY model, are $\eta=0.04$ and $\nu=0.67$, so $(1-\eta)\nu=\gamma_s=0.64$.  In three and higher
dimensions for the Bose Hubbard model, the critical exponents are mean-field like, with $\eta=0$ and $\nu=0.5$, so
$(1-\eta)\nu=\gamma_s=0.5$.  The one-dimensional case has Kosterlitz-Thouless behavior~\cite{kosterlitz_thouless}, where $\eta=0.25$, and the
divergence of the correlation length has a Kosterlitz-Thouless exponential form $\xi\propto \exp[W/\sqrt{x_c-x}]$,
with $x_c$ the critical point for the Mott insulator to superfluid transition.

This critical scaling behavior does not provide enough information for us to determine an ansatz for the momentum distribution function over all momentum, because the distribution function is not critical for nonzero momentum. We use the exact solution in the infinite-dimensional limit, as given by the RPA solution, to guide us in how to proceed to develop an appropriate scaling ansatz.   The RPA form for the momentum distribution function, as discussed above, and reexpressed in terms of $\xi_{\bf k}$ and $x$, is given by~\cite{rpa}
\begin{equation}
 n_{\bf k}=-\frac{1}{2}+\frac{n_{}+\frac{1}{2}+\xi_{\bf k}x}{\sqrt{1+4(2n_{}+1)\xi_{\bf k}x+4\xi_{\bf k}^2x^2}},
\label{eq: rpa}
\end{equation}
and this is the exact solution in infinite dimensions.  A quick examination of the strong-coupling expansion for
arbitrary dimensions, shows that the $O(1)$ terms are the same for all dimensions, when expressed in terms of
$x$ and $\xi_{\bf k}$, and it is only the $1/d^n$ corrections that differ for the different dimensions.  Hence, the  power-series expansion of the RPA form must produce all of the $O(1)$ terms.  In finite dimensions, only $1/d^n$
corrections are allowed.  This motivates the following scaling ansatz for the momentum distribution function
in two or higher dimensions (on a bipartite lattice)
\begin{equation}
 n_{\bf k}=-\frac{1}{2}\label{eq: scaling_ansatz}
+\frac{n_{}+\frac{1}{2}+\xi_{\bf k}x+\frac{c^\prime}{d^2} x^2+2\frac{e^\prime}{d^2} \xi_{\bf k}x^3}
{[1+2\bar a\xi_{\bf k}x+4\bar b\xi_{\bf k}^2x^2+\frac{\bar c}{d^2} x^2+8\bar d \xi_{\bf k}^3x^3+2\frac{\bar e}{d^2}\xi_{\bf k}x^3]^{\gamma_s}},
\end{equation}
with $d$ the spatial dimension.
Note that in three dimensions, since $\gamma_s=0.5$ which is the same power law as in infinite dimensions, we must have $\bar d=0$.  We will see this occur in the analysis below.

In order to determine the parameters in the scaling ansatz, we propose three requirements of the formula in Eq.~(\ref{eq: scaling_ansatz}): (i) the power-series expansion of the scaling ansatz, in powers of $x$, must reproduce the strong-coupling expansion through the given order (in our case through third order) as shown in Eq.~(\ref{eq: third_order_general}); (ii) we choose the scaling form to have the exact critical point $x_c$, as determined by QMC, DMRG, or scaling results of a strong-coupling expansion for the phase diagram; and (iii) we require the integral of $n_{\bf k}$ over all momentum to give $n_{}$, the density of the bosons in the Mott phase.  In two and higher dimensions, these three requirements will determine all of the parameters, which we now show; in one-dimension, we use additional information to determine the Kosterlitz-Thouless constant $W$, which then allows us to determine the complete scaling
form.

We begin with the infinite-dimensional case where the hopping scales like $1/d$ so that $x$ is finite, but the coefficients $\bar c$, $c^\prime$, $\bar e$, and $e^\prime$ all vanish because they are $1/d^2$ corrections. Expanding the scaling ansatz in a power series in $x$ and equating the coefficients of the powers of $x$ with the strong-coupling expansion in Eq.~(\ref{eq: third_order_general}), yields the following: $\bar a=(2n_{}+1)/\gamma_s=2(2n_{}+1)$, $\bar b=1/2\gamma_s=1$, and $\bar d=0$, so we recover the RPA result in Eq.~(\ref{eq: rpa}). Since the critical behavior occurs at the point where $1+4(2n_{}+1)\xi_{\bf k}x_c+4\xi_{\bf k}x_c^2=0$, and we evaluate for ${\bf k}=0$, where $\xi_{\bf 0}=-1$, we immediately find that $x_c=(n_{}+1/2)-\sqrt{n_{}(n_{}+1)}$, which is the exact critical point~\cite{fisher_etal,strong2} for all $n_{}$. Hence, one can see that this approach automatically produces the right behavior for the large-dimensional limit.

Note that the curvature of the RPA momentum distribution, with respect to $\xi_{\bf k}$, is always one sign.  In the truncated third-order strong-coupling expansion, the curvature of the momentum distribution function changes sign at $\xi_{\bf k}=1$ when $x\approx 0.034$.  This effect occurs for all finite dimensions as well.  The scaled results, that are shown below, do not have a change in the sign of the curvature, and we expect that this does not occur in any of the exact solutions of the Bose Hubbard model.

Since the momentum distribution function depends on the correlation length at ${\bf k}=0$, and so does the phase diagram, it is interesting to try a phenomenological exercise, where we take the critical behavior determined via our scaling approach for the momentum distribution and relate it to a determination of the phase diagram.  Since we have a power law of a polynomial, instead of the simplest scaling dependence, which would go like $\sqrt{x_c-x}$, such an approach is similar to summing an infinite number of terms in the expansion for the Mott phase lobes in the phase diagram.  As an example, we make the following scaling ansatz for the Mott lobes
\begin{equation}
 \left .\frac{\mu}{U}\right |_\pm =n_{}+A(x)\pm B(x) [{\rm scaling~polynomial}]^{Z\nu},
\label{eq: lobe_scale_ansatz}
\end{equation}
where the scaling polynomial is the polynomial used for the momentum distribution at ${\bf k}=0$ [which is $1-4(2n_{}+1)x+4x^2$ for the infinite-dimensional case], and $A(x)$ and $B(x)$ are polynomials in $x$. Fitting the parameters to the third-order expansion for the Mott phase lobes, we find for the infinite-dimensional case that
\begin{equation}
 \left .\frac{\mu}{U}\right |_\pm =n_{}-\frac{1}{2}-x\pm \frac{1}{2}\sqrt{1-4(2n_{}+1)x+4x^2},
\label{eq: infd_phase_lobe}
\end{equation}
which is the exact solution~\cite{fisher_etal,strong2}.

\begin{figure}[th]
\centerline{\includegraphics [width=3.3in, angle=0, clip=on]  {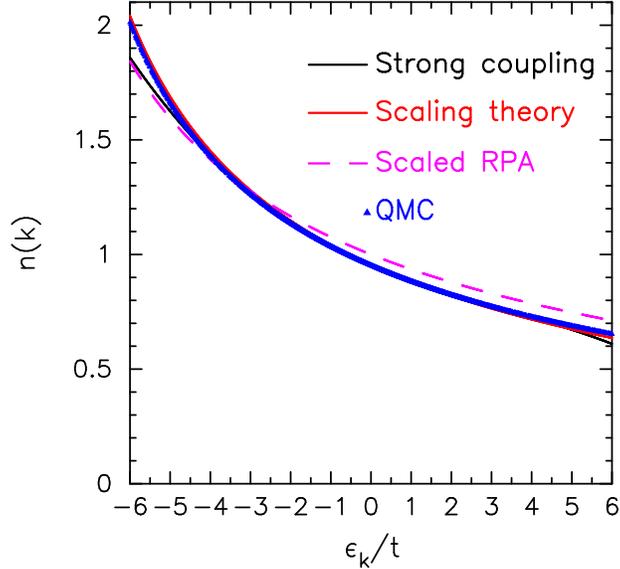}}

\caption[]{
(Color online)  Momentum distribution function for the three-dimensional case with $x=0.0625$ as a function of the
band energy $\epsilon_{\bf k}$.  Note how the QMC data agrees better with the scaling theory results than it does with
the strong-coupling results or the scaled RPA, although deviations can be seen in the data.
}
\label{fig: 3d_x=0.0625}
\end{figure}

In finite dimensions, we will consider only the $n_{}=1$ case, because good numerical data is available for the Mott phase boundary in one, two, and three dimensions and we want to ensure that we produce the correct critical point $x_c$.
We start with the three-dimensional case.  Taking $n_{}=1$, we expand Eq.~(\ref{eq: scaling_ansatz}) in a power series in $x$, and compare with the strong-coupling expansion in Eq.~(\ref{eq: 3d}).  We find $\bar a=6$, $\bar b=1$, $\bar c=144+4c^\prime/3$, $\bar d=0$, and $\bar e=224/3+58c^\prime/9+4e^\prime/3$.  At this point, the constants $c^\prime$ and $e^\prime$ are not determined.  We fix $e^\prime$, by requiring the momentum distribution at ${\bf k}=0$ to diverge at the critical point $x_c=0.10224$ as determined by QMC simulation~\cite{prokofiev_3d}. This produces the equation
\begin{equation}
 1-12x_c+\left ( 20+\frac{4}{27}c^\prime\right )x_c^2-\left ( \frac{448}{27}+\frac{116}{81}c^\prime+\frac{8}{27}e^\prime\right )x_c^3=0.
\label{eq: 3d_crit}
\end{equation}
Setting $x_c=0.10224$, and solving for $e^\prime$, yields
\begin{equation}
 e^\prime=-122.2743+0.0571205c^\prime.
\label{eq: 3d_eprime}
\end{equation}
(Note that if we instead set the coefficients $c^\prime$ and $e^\prime$ to zero, then the critical point would lie at $x_c=0.09805$, which is about a 4.3\% error.)
The coefficient $c^\prime$ is determined by requiring $\int d^3k n_{\bf k}=1$; we find that $c^\prime$ ranges from 0 at $x=0$ out to $c^\prime=-1.86$ as $x\rightarrow x_c$. A simple polynomial fit to the behavior of $c^\prime(x)$ is
\begin{equation}
 c^\prime(x)=0.017166-0.71982x-161.093x^2-109.614x^3.
\label{eq: 3d_cprime}
\end{equation}

We compare the strong-coupling perturbation theory to numerically exact
results performed with
world-line quantum Monte Carlo simulations of the Bose Hubbard model
that employ the  directed-loop algorithm~\cite{sandvick},
in particular, its continuous-imaginary-time variant~\cite{kawashima}.
We have further improved the algorithm by omitting one-site vertices
corresponding to the U-term~\cite{kato} and also two-body vertices
corresponding to the hopping term~\cite{kato2}. The latter modification is
useful in reducing the memory and was crucial in the present
simulation of the largest system ($L=64$). The accuracy of the
method is tested by comparing with exact diagonalization for
small systems, and verifies the critical exponents with known results for
the $d+1$-dimensional XY model. To further test that the true equilibrium distribution
is sampled on large systems, several independent runs with varying
lattice sizes are carried out, showing no systematic deviation, thereby
ensuring that our numerical results are ``exact" except for
statistical errors. This QMC approach has already been applied to the problem of determining how the momentum distribution changes when the system becomes superfluid~\cite{nandini2}.

We also compare the momentum distribution to RPA results.  Since the RPA has a critical value of $x$ that is smaller than the true critical value in finite dimensions, we plot the RPA results in Eq.~(\ref{eq: rpa}) at a rescaled hopping value, corresponding to the same fractional amount of $x_c$.  Namely, we choose $x_{RPA}=0.0857864x/x_c(d)$. We call this the scaled RPA momentum distribution.

\begin{figure}[th]
\centerline{\includegraphics [width=3.3in, angle=0, clip=on]  {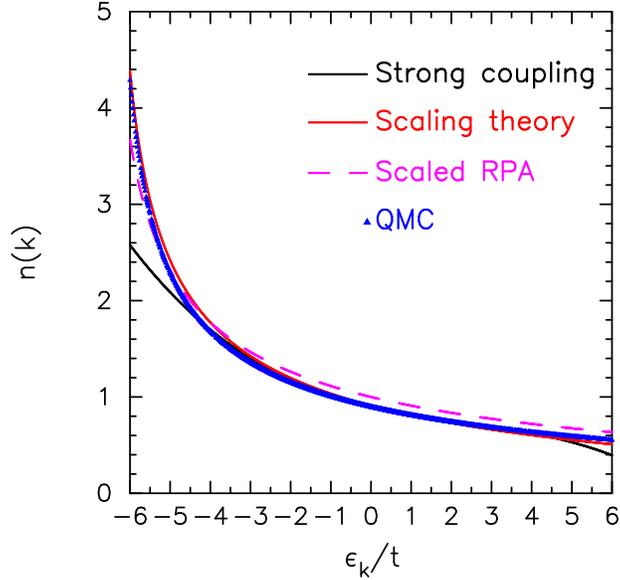}}
\caption[]{
(Color online)  Momentum distribution function for the three-dimensional case with $x=0.09$ as a function of the
band energy $\epsilon_{\bf k}$.  Once again, the QMC data agrees better with the scaling theory than it does with
the strong-coupling results or the scaled RPA, although deviations can still be seen in the data.  Note that the scaled RPA works better than the truncated strong-coupling expansion.
}
\label{fig: 3d_x=0.09}
\end{figure}

The scaled results of the strong-coupling perturbation theory
fit the numerical QMC data quite well.  We compare with data at $x=0.0625$ and $x=0.09$ in Figs.~\ref{fig: 3d_x=0.0625} and \ref{fig: 3d_x=0.09}, respectively. The QMC data is for a $48\times 48\times 48$ lattice at a temperature $T=0.1t$ ($T=0.025t$ for $x=0.09$); in all cases, we have carefully checked that the finite-size effects and the finite-temperature effects are much smaller than the symbol size in all of our results. Note how the QMC data follows the scaled curve much better than the strong-coupling curve, although there are definitely differences between the two. The deviations between the QMC data and the scaling result are real and larger than the finite-size or finite-temperature effects.  This simply reflects the fact that the scaling result is not an exact interpolation formula for the momentum distribution. As expected, the momentum distribution is peaked at zero momentum, and as one approaches the critical point at $x=0.10224$, the peak becomes sharper. One can also see that the truncated third-order expansion is not too accurate.  As we already mentioned above, the curvature for momenta near the zone boundary has the wrong sign even for quite small hopping.  It also underestimates the size of the peak at zero momentum, and this gets worse as we approach the critical point.  Nevertheless, the strong-coupling expansion is quite accurate for small enough hopping, and the fact that it agrees essentially exactly with both the scaled results and the QMC simulations, provides an independent check that all of these different approaches are working to high precision.

\begin{figure}[th]
\centerline{\includegraphics [width=3.3in, angle=0, clip=on]  {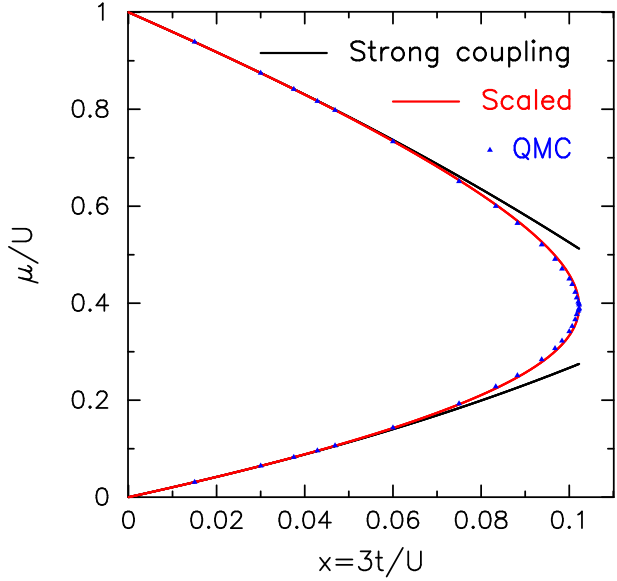}}

\caption[]{
(Color online)  Phase diagram of the three-dimensional Bose Hubbard model.  Note how the truncated strong-coupling expansion does not agree so well with the QMC data~\cite{prokofiev_3d}, but the scaled results nearly fit the Mott phase lobe perfectly.
}
\label{fig: 3d_phase}
\end{figure}

We now try the phenomenological approach on the three-dimensional phase diagram.  Here we have some uncertainty in how to proceed, because the scaling polynomial has freedom in our ability to vary the $c^\prime$ coefficient.  We can either modify the scaling polynomial to represent the changes in $c^\prime$, or we can fix $c^\prime$ at a specific value and proceed from there.  It turns out that we get better results if we fix $c^\prime=0$ when calculating the phase diagram (especially for the two-dimensional case below). So we adopt that as our procedure (note we do not also set $e^\prime=0$, because that would produce the wrong critical point for this phenomenological approach).  The result for the Mott phase lobes is
\begin{equation}
 \left . \frac{\mu}{U}\right |_\pm=\frac{1}{2}-x-\frac{1}{2}x^2+x^3\pm\frac{\frac{1}{2}-\frac{1}{2}x^2-8.81514x^3}{\sqrt{1-12x+20x^2+16.67387x^3}}.
\label{eq: 3d_mott_lobes}
\end{equation}
These results are plotted versus the QMC calculations~\cite{prokofiev_3d} in Fig.~\ref{fig: 3d_phase}.  One can see that while the truncated strong coupling expansion~\cite{strong1,strong2} does not agree so well with the QMC data near the critical point, the agreement of the scaled curves is excellent.

\begin{figure}[th]
\centerline{\includegraphics [width=3.3in, angle=0, clip=on]  {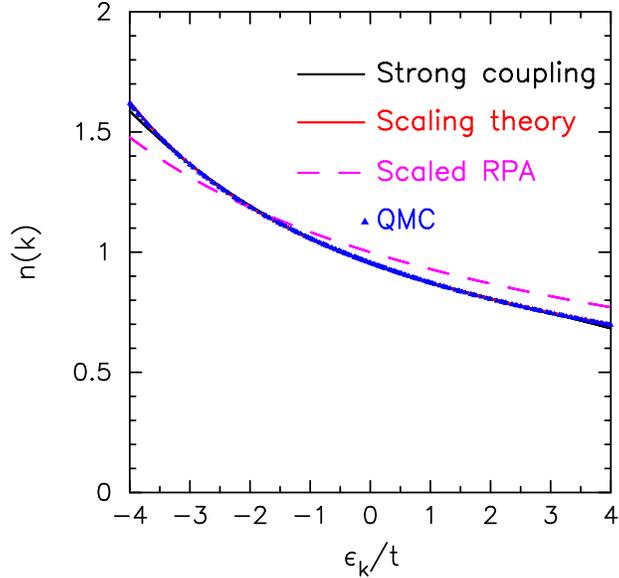}}

\caption[]{
(Color online)  Momentum distribution function in two dimensions with $x=0.05$.  We plot the strong-coupling expansion against the scaling theory results, the scaled RPA, and QMC simulations.  Note how the QMC results agree much better with the scaled results
and do not show the change in curvature near $\xi_{\bf k}=1$. In addition, the scaled RPA doesn't work as well here as it did in three dimensions.
}
\label{fig: 2d_x=0.05}
\end{figure}

Next, we move on to two dimensions.  Recall that $\gamma_s=0.64$ in this case. Going through the same procedure outlined above produces the following solution for the coefficients in the scaling polynomial: $\bar a=3/\gamma_s=4.6875$;
$\bar b=-17/2\gamma_s+9/2\gamma_s^2=-2.29492$; $\bar c=48/\gamma_s+2c^\prime/3\gamma_s=75.0+1.04167c^\prime$; $\bar d=33/\gamma_s-51/2\gamma_s^2+9/2\gamma_s^3=6.47278$; and $\bar e=-256/\gamma_s+144/\gamma_s^2-2c^\prime/9\gamma_s+2c^\prime/\gamma_s^2+2e^\prime/3\gamma_s=-48.4375+4.53559c^\prime+1.04167e^\prime$.  Once again, $c^\prime$ and $e^\prime$ are as yet undetermined.  We find $e^\prime$ by requiring the critical point at ${\bf k}=0$ to occur at the QMC and strong-coupling critical point $x_c=0.11948$~\cite{prokofiev_2d}.  The critical point is found when
\begin{eqnarray}
 1&-&9.375x_c+(9.57032+0.260418c^\prime)x_c^2\label{eq: 2d_crit}\\
&+&(-27.56349-2.26780c^\prime-0.260418e^\prime)x_c^3=0.
\nonumber
\end{eqnarray}
(If we set $c^\prime=e^\prime=0$, then the critical point would lie at $x_c=0.11579$ which is a 3.2\% error.) Substituting in $x_c=0.11948$, then yields $e^\prime=-68.7054-0.338706c^\prime$.  The parameter $c^\prime$ is then determined by requiring the integral of $n_{\bf k}$ over all momentum to equal one.  We find that $c^\prime$ ranges from approximately $-115$ at $x= 0$ to $c^\prime\approx -224$ at $x=0.119$, but for values of $x$ larger than about $0.1169$, there is no value of $c^\prime$ that gives the total particle density to be exactly one---the error is about $1.5\%$ at $x=0.119$ when we choose the best fit $c^\prime$. A simple fit of $c^\prime(x)$ is
\begin{equation}
 c^\prime(x)=-99-13.7(1-7.914x)^{-0.77}.
\label{eq: 2d_cprime_fit}
\end{equation}

\begin{figure}[th]
\centerline{\includegraphics [width=3.3in, angle=0, clip=on]  {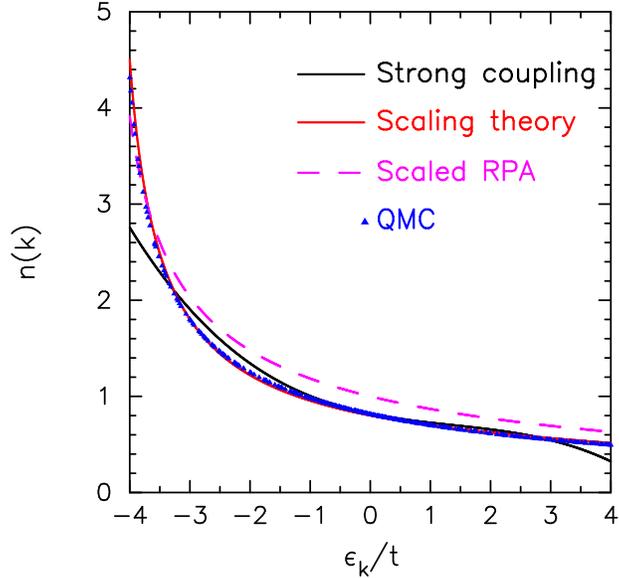}}

\caption[]{
(Color online)  Momentum distribution function in two dimensions with $x=0.1$, which is close to the critical point.  We plot the strong-coupling expansion against the scaling theory results, the scaled RPA, and the QMC simulations.  Note how the QMC results agree much better with the scaling theory results.
}
\label{fig: 2d_x=0.1}
\end{figure}

We compare our analytic expressions to QMC data in two dimensions on a $48\times 48$ lattice with $T=0.05$.  In Fig.~\ref{fig: 2d_x=0.05}, we plot a case far from the critical point with $x=0.05$.  The scaling curve and the truncated strong-coupling expansion are both quite close to each other here, but one can see how the curvature has changed in the strong-coupling expansion but not in the data nor in the scaled curve.  One also can see systematically that the QMC data agrees better with the scaled curve than the strong-coupling expansion.  Moving on to a point much closer to the critical point at $x=0.1$, we show the same plots in Fig.~\ref{fig: 2d_x=0.1}  with the QMC data on a $48\times 48$ lattice with $T=0.00625$.  Here, one can see a much more dramatic difference between the truncated strong-coupling results and the scaled results.  While there definitely are some minor discrepancies with the QMC data and the scaled results, the agreement is, in general, outstanding.  Note that we plot the momentum distribution versus $\epsilon_{\bf k}$ instead of {\bf k}, because in the strong coupling expansion all momentum dependence is summarized in $\epsilon_{\bf k}$ through third order, so there is limited other momentum dependence.  For the QMC data, we average the small number of degenerate energy values.

\begin{figure}[th]
\centerline{\includegraphics [width=3.3in, angle=0, clip=on]  {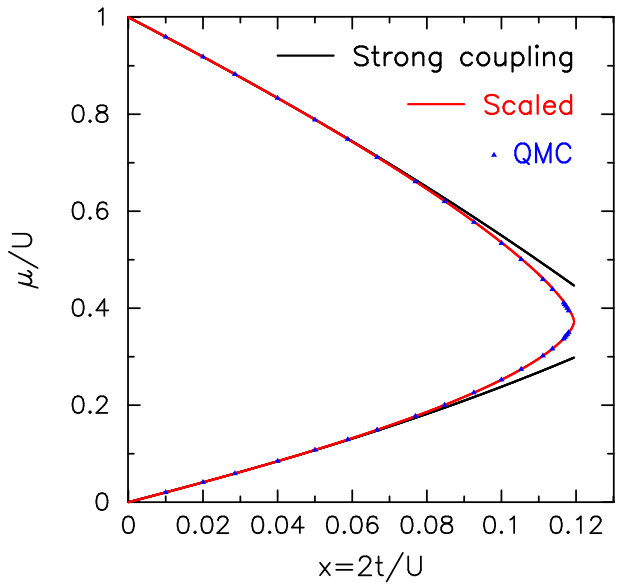}}

\caption[]{
(Color online)  Phase diagram of the two-dimensional Bose Hubbard model.  Note how the truncated strong-coupling expansion~\cite{strong1,strong2} does not agree so well with the QMC data~\cite{prokofiev_2d}, but the scaled results nearly fit the Mott phase lobe perfectly.
}
\label{fig: 2d_phase}
\end{figure}

We finally try the phenomenological fit to the phase diagram by using the scaling polynomial in the power law and forcing the third-order strong coupling expansion to agree with the phenomenological scaling ansatz.  Once again, we set $c^\prime=0$ when we do this, because the agreement is significantly worse with different $c^\prime$ values.  Because $c^\prime$ assumes much larger values in two dimensions in order to get the right integrated weight in the momentum distribution, this is a significant assumption we are making, but as seen in the final results, the assumption seems reasonable because the agreement is quite good.

Following an identical procedure to what was done in the three-dimensional case (with $c^\prime$ set equal to zero), we find
\begin{eqnarray}
 \left . \frac{\mu}{U}\right |_\pm&=&\frac{1}{2}-x-\frac{3}{4}x^2+\frac{3}{2}x^3\label{eq: 2d_mott_lobes}\\
&\pm&\frac{\frac{1}{2}+0.14063x-0.21460x^2-3.87043x^3}{[1-9.375x+9.5704x^2+9.6757x^3]^{0.67}}.
\nonumber
\end{eqnarray}
These results are plotted versus the QMC calculations~\cite{prokofiev_2d} in Fig.~\ref{fig: 2d_phase}.  Once again note that while the truncated strong-coupling expansion does not agree so well with the QMC data near the critical point,  the scaled curves lie essentially on top of the QMC data.

The one-dimensional case is different from higher dimensions because the scaling behavior is not power law, but
instead is the Kosterlitz-Thouless form of the two dimensional XY model.  Hence, we modify our scaling ansatz to
\begin{eqnarray}
n_{\bf k}&=&-\frac{1}{2}
+\left [n_{}+\frac{1}{2}+\xi_{\bf k}x+c^\prime x^2+2e^\prime \xi_{\bf k}x^3\right ]\nonumber\\
&\times&
\exp\left [ -W^\prime+\frac{W^\prime}{\sqrt{
1+2\bar a\xi_{\bf k}x+4\bar b\xi_{\bf k}^2x^2+\bar c x^2+8\bar d \xi_{\bf k}^3x^3+2\bar e \xi_{\bf k}x^3}}\right ],
\label{eq: scaling_ansatz_1d}
\end{eqnarray}
which replaces the power law divergence by the appropriate exponential divergence. Because the exponent $\eta=0.25$ for
the two-dimensional XY model, we have that $W^\prime=0.75 W$, with $W$ the parameter in the Kosterlitz-Thouless fit
to the one-dimensional Mott phase diagram.  Using the data of Elstner and Monien~\cite{elstner2}, we fit the gap function $\Delta(x)$ to the Kosterlitz-Thouless form
\begin{equation}
\left [ \ln \Delta(x)\right ]^2\label{eq: kt_fit}
=\frac{A+Bx+Cx^2+Dx^3+Ex^4+Fx^5+Gx^6}{1+Hx+Ix^2+Jx^3+Kx^4+Lx^5+Mx^6+Nx^7},
\nonumber
\end{equation}
by using a Pade approximant for the pole that develops in the square of the logarithm of the gap function.
Note that one needs to do the Pade approximant for the square of the logarithm of the power series in order to obtain a robust fit [instead of doing a series or Pade approximation for $\Delta(x)$ first and then taking the square of the logarithm of the resulting series or Pade approximant].  The critical point is $x_c=0.29981$ and the parameter $W$ becomes $W=1.7241$ or $W^\prime=1.2931$.

\begin{figure}[thb]
\centerline{\includegraphics [width=3.3in, angle=0, clip=on]  {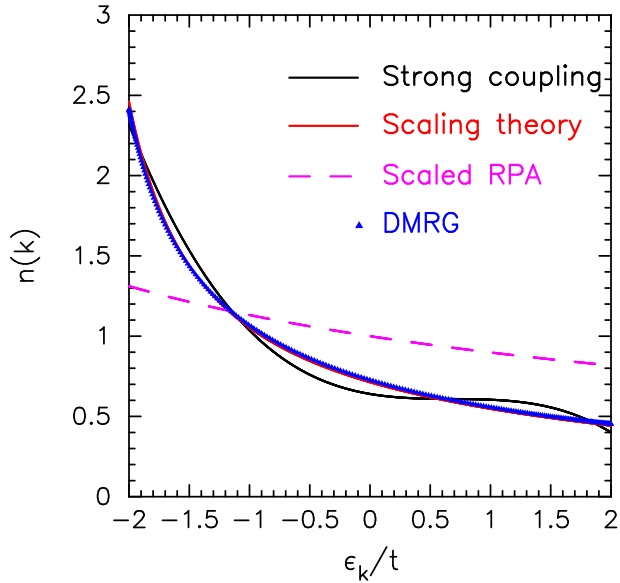}}
\caption[]{
(Color online)  Momentum distribution function in one dimension with $x=0.1$, which is far from the critical point.  We plot the strong-coupling expansion against the scaling theory results, the scaled RPA, and the DMRG calculations.  Note how the DMRG results agree much better with the scaling theory results than the truncated expansion or the scaled RPA.
}
\label{fig: 1d_x=0.1}
\end{figure}

Now we solve for the coefficients in the scaling form just as we did in higher dimensions.  First we ensure that the power-series expansion of the scaling form reproduces the strong-coupling expansion through the third order in $x$,
then we ensure that the denominator of the square root in the exponential diverges at $x_c$.  These two conditions yield
$\bar a=4.6400$, $\bar b=3.0006$, $\bar c =37.1201+1.0311 c^\prime$, $\bar d=9.4879$, $\bar e=64.2632+6.8329c^\prime$,
and $e^\prime=-9.4630-0.3190c^\prime$.  The coefficient $c^\prime$ is adjusted to guarantee that the integral of the
momentum distribution over all momentum is equal to one.  We find that
$c^\prime\approx -7.92-15.16x$ in order to satisfy the sum rule.

\begin{figure}[thb]
\centerline{\includegraphics [width=3.3in, angle=0, clip=on]  {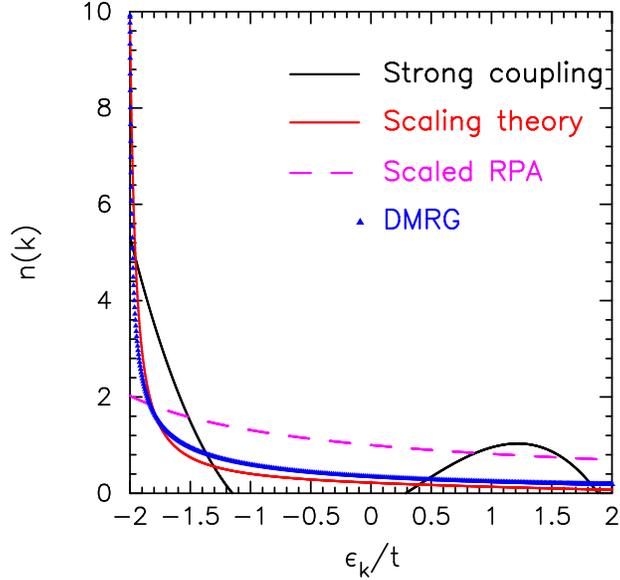}}
\caption[]{
(Color online)  Momentum distribution function in one dimension with $x=0.2$, which is two-thirds of the way to the critical point.  We plot the strong-coupling expansion against the scaling theory results, the scaled RPA, and the DMRG calculations.  Note how the DMRG results agree much better with the scaling theory results than the truncated expansion or the scaled RPA, but one can see that the scaling approach is beginning to fail.
}
\label{fig: 1d_x=0.2}
\end{figure}

We compare the scaled strong-coupling perturbation theory to the numerical calculations in one dimension from the density matrix renormalization group (DMRG) approach~\cite{kollath} (provided to us by C. Kollath).  Those calculations are essentially exact except for finite-size effects which become more important as we approach the critical point at $x=0.29981$.  In Fig.~\ref{fig: 1d_x=0.1}, we compare the different approximations to the DMRG calculations.  One can immediately see that although the truncated expansion has a nonmonotonic dependence on $\epsilon_{\bf k}$, the scaled approach essentially agrees exactly with the DMRG calculations.

Next, we compare the different approximate results to the DMRG calculations for $x=0.2$ in Fig.~\ref{fig: 1d_x=0.2}.  Here we see that while the scaled results still agree well with the DMRG results near ${\bf k}=0$, the agreement is not so good throughout the Brillouin zone, and it is clear that the approximation is becoming inadequate. When we compare results for large values of the hopping, such as $x=0.25$, the scaled results become negative over about half of the Brillouin zone, which is unphysical.

We do not go through the phenomenological exercise of comparing our results to the phase diagram in the one-dimensional case as we did previously for higher dimensions.  This is primarily because we could see the approximate scaled results were breaking down around $x\approx 0.2$, so it is unlikely that a phenomenological approach for the phase diagram would be accurate in this case.  In general, the strong coupling approach is more accurate in higher rather than lower dimensions.

\section{Conclusions}

In this work, we have shown how one can generalize strong coupling perturbation theory from an expansion for the many-body energy levels, or for different ground-state correlation functions, to a direct expansion for the many-body Green's function at finite temperature.  Here, we focused on applying the expansion to the problem of determining the momentum distribution in the bulk for the Bose Hubbard model within the Mott-insulating phase.  By applying a scaling ansatz, that was motivated by recent work on the RPA, we are able to find accurate analytic expressions for the momentum distributions that hold nearly up to the critical point in two and three dimensions (the results for one-dimension are not quite as good).  In addition, we showed how one can apply the results for the momentum distribution function to create a phenomenological theory for the Mott phase lobes.  Comparing these results to QMC simulations showed excellent agreement in two and three dimensions.

The strong coupling formalism as developed here can be used, as we have indicated, to obtain a strong-coupling expansion for the self-energy, and to include inhomogeneous features like a harmonic trap or disorder potential, and the effects of thermal excitations. It can also be readily adapted to nonequilibrium cases such as moving the origin of the trap or modulating the optical lattice depth for Bragg spectroscopy.  The quantum Monte Carlo approach can be generalized to calculate dispersion relations, densities of states, and real time dynamics. We intend to examine those problems in the future.

\acknowledgments
J. K. F. and H. R. K. acknowledge support under ARO Grant W911NF0710576 with funds from the DARPA OLE Program. H.R.K also acknowledges support from DST (India) as a J. C. Bose Fellow. Part of this work was completed during a stay at the Aspen Center for Physics. N. T. acknowledges support under ARO grant number W911NF0810338 with funds from the DARPA OLE Program. The quantum Monte carlo simulations were carried out at the Supercomputer Center, Institute for Solid State Physics, University of Tokyo. Y. K. and N. K. acknowledge support under MEXT Grant-in-Aid for Scientific Research (B) (No. 19340109) and a Grant-in-Aid for JSPS Fellows 
We also acknowledge useful discussions with
D. Arovas,
A. Auerbach,
C. Kollath,
H. Monien,
W. Phillips,
J. Porto,
R. R. P. Singh
and I. Spielmann.
DMRG data for the momentum distribution in one dimension were provided by C. Kollath.
The QMC data for the phase diagram in two and three dimensions were provided by B. Capogrosso-Sansone.

\appendix

\section{Imaginary time integrals needed for the strong-coupling expansion
\label{appendixa}}

Consider the i-time ordered integral
\bea
&&I_m(\beta; \mathcal{E}_{\alpha_0}, \mathcal{E}_{\alpha_m}, \cdots, \mathcal{E}_{\alpha_1}) \equiv \frac {e^{-\beta\mathcal{E}_{\alpha_0}}} {\mathfrak{Z}}  \int_0^{\beta} d\tau_m \int_0^{\tau_m} d\tau_{m-1} \cdots \int_0^{\tau_2} d\tau_{1} \non \\
&\times& \; e^{[\tau_m(\mathcal{E}_{\alpha_0}-\mathcal{E}_{\alpha_m}) + \tau_{m-1} (\mathcal{E}_{\alpha_m}-\mathcal{E}_{\alpha_{m-1}}) + \cdots + \tau_2 (\mathcal{E}_{\alpha_3}-\mathcal{E}_{\alpha_2}) + \tau_1 (\mathcal{E}_{\alpha_2} -\mathcal{E}_{\alpha_1}) ]}.
\eea
It is easy to see that the sequence of functions $I_m$ satisfy the recursion relation:
\beq
I_m(\tau; \mathcal{E}_{\alpha_0}, \mathcal{E}_{\alpha_m}, \cdots, \mathcal{E}_{\alpha_1}) = \int_0^{\tau} d\tau^\prime e^{-(\tau - \tau^\prime)\mathcal{E}_{\alpha_0}} I_{m-1}(\tau^\prime; \mathcal{E}_{\alpha_m}, \mathcal{E}_{\alpha_{m-1}} \cdots, \mathcal{E}_{\alpha_1}).
\eeq
Taking the Laplace transform of both sides, it is straightforward to see that
\bea
\mathcal{L}[ I_m(\tau; \mathcal{E}_{\alpha_0}, \mathcal{E}_{\alpha_m}, \cdots, \mathcal{E}_{\alpha_1}); s] & \equiv &
\int_0^{\infty} d\tau e^{- s \tau} I_m(\tau; \mathcal{E}_{\alpha_0}, \mathcal{E}_{\alpha_m}, \cdots, \mathcal{E}_{\alpha_1}) \non \\
& = & \int_0^{\infty} d\tau \int_0^{\tau} d\tau^\prime e^{- s \tau}  e^{-(\tau - \tau^\prime)\mathcal{E}_{\alpha_0}} I_{m-1}(\tau^\prime; \mathcal{E}_{\alpha_m}, \mathcal{E}_{\alpha_{m-1}} \cdots, \mathcal{E}_{\alpha_1}) \non \\
& = & \int_0^{\infty} d\tau^\prime \int_{\tau^\prime}^{\infty}  d\tau e^{- (s + \mathcal{E}_{\alpha_0}) \tau}  e^{\tau^\prime \mathcal{E}_{\alpha_0}} I_{m-1}(\tau^\prime; \mathcal{E}_{\alpha_m}, \mathcal{E}_{\alpha_{m-1}} \cdots, \mathcal{E}_{\alpha_1}) \non \\
& = & \int_0^{\infty} d\tau^\prime \frac{e^{- s \tau^\prime}}{s + \mathcal{E}_{\alpha_0}} I_{m-1}(\tau^\prime; \mathcal{E}_{\alpha_m}, \mathcal{E}_{\alpha_{m-1}} \cdots, \mathcal{E}_{\alpha_1}) \non \\
& = & \frac{1}{s + \mathcal{E}_{\alpha_0}} \mathcal{L}[I_{m-1}(\tau; \mathcal{E}_{\alpha_m}, \mathcal{E}_{\alpha_{m-1}} \cdots, \mathcal{E}_{\alpha_1}) ; s].
\eea
Iterating this, and noting that $I_0(\tau; \mathcal{E}_{\alpha_1}) = e^{-\tau \mathcal{E}_{\alpha_1}} / \mathfrak{Z}$, which implies that $ \mathcal{L}[I_0(\tau; \mathcal{E}_{\alpha_1});s] = [ \mathfrak{Z} (s + \mathcal{E}_{\alpha_1})]^{-1}$, we find
\beq
\mathcal{L}[ I_m(\tau; \mathcal{E}_{\alpha_0}, \mathcal{E}_{\alpha_m}, \cdots, \mathcal{E}_{\alpha_1}); s] = \frac {1}{\mathfrak{Z}} \prod_{\ell = 0, m} \frac{1}{(s + \mathcal{E}_{\alpha_\ell})}.
\eeq
Taking the inverse Laplace transform yields
\beq
I_m(\tau; \mathcal{E}_{\alpha_0}, \mathcal{E}_{\alpha_m}, \cdots, \mathcal{E}_{\alpha_1}) = \int_{\gamma - i\infty}^{\gamma + i\infty} \frac{ds}{2 \pi i} \; \frac {e^{\tau s}}{\mathfrak{Z}} \prod_{\ell = 0, m} \frac{1}{(s + \mathcal{E}_{\alpha_\ell})},
\label{eq: I_m-inv-LT}
\eeq
with $\gamma > \max{(\mathcal{E}_{\alpha_0}, \mathcal{E}_{\alpha_m}, \cdots, \mathcal{E}_{\alpha_1})}$, so that all the singularities of the integrand lie to the left of the integration contour in the complex $s$-plane. The integral is straightforwardly evaluated using the contour integration techniques. When all the energies $ \mathcal{E}_{\alpha_0}, \mathcal{E}_{\alpha_m}, \cdots, \mathcal{E}_{\alpha_1} $ are distinct, we get one contribution from each of the $m+1$ simple poles of the integrand in Eq.~(\ref{eq: I_m-inv-LT}), leading to Eq.~(\ref{eq: I_m-0}). If  one and only one pair of energies are equal, say, $\mathcal{E}_{\alpha_r} = \mathcal{E}_{\alpha_p}$, then the integrand of Eq.~(\ref{eq: I_m-inv-LT}) has $m-1$ simple poles and one double pole, and we get Eq.~(\ref{eq: I_m-1}). One can similarly extend the results to other cases, corresponding to two double poles, or one triple pole, {\it etc.}

\section{Final results for the third-order expansion terms\label{appendixb}}

Explicit forms for the third-order coefficients in the strong-coupling expansion are presented here (for brevity only in the second form, as discussed in Sec.~II):
\bea
&&\mathfrak{C}^{(3)}_{j j_2 j_1 j^\prime} =  \mathfrak{t}_{j j_2} \mathfrak{t}_{j_2 j_1} \mathfrak{t}_{j_1 j^\prime} \, \sum_{n, n_1, n_2, n^\prime}  \; \rho_{j, n} \,\rho_{j_1, n_1} \, \rho_{j_2, n_2} \, \rho_{j^\prime, n^\prime}  \; \times \non \\
& \{ & [ \frac{ n (n_2 + 1)(n_1 + 1)(n^\prime + 1)}{(\epsilon^-_{j, n} +  \epsilon^+_{j_2, n_2})(\epsilon^-_{j, n} +  \epsilon^+_{j_1, n_1})(\epsilon^-_{j, n} + \epsilon^+_{j^\prime, n_j^\prime})}  +  \frac{ (n + 1)(n_2 + 1) n_1 (n^\prime + 1)} {(\epsilon^+_{j, n} +  \epsilon^-_{j_1, n_1})(\epsilon^-_{j_1, n_1} +  \epsilon^+_{j_2, n_2})(\epsilon^-_{j_1, n_1} +  \epsilon^+_{j^\prime, n^\prime })} \non \\
& + &   \frac{ (n + 1) (n_2 + 1)(n_1 + 1) n^\prime } {(\epsilon^+_{j, n} +  \epsilon^-_{j^\prime, n^\prime}) (\epsilon^+_{j_2, n_2} +  \epsilon^-_{j^\prime, n^\prime})(\epsilon^+_{j_1, n_1} +  \epsilon^-_{j^\prime, n^\prime})} + \frac{ (n + 1) n_2 (n_1 + 1)  (n^\prime + 1)} {(\epsilon^+_{j, n} +  \epsilon^-_{j_2, n_2})(\epsilon^+_{j_1, n_1} +  \epsilon^-_{j_2, n_2})(\epsilon^-_{j_2, n_2} +  \epsilon^+_{j^\prime, n^\prime })} ] \non \\
& + & [ \frac{n n_2 (n_1 + 1) (n^\prime + 1)}{(\epsilon^-_{j, n} +  \epsilon^+_{j^\prime, n^\prime})(\epsilon^-_{j, n} +  \epsilon^+_{j_1, n_1})(\epsilon^-_{j_2, n_2} +  \epsilon^+_{j_1, n_1})} + \frac{(n + 1) n_2 n_1 (n^\prime + 1)}{(\epsilon^+_{j, n} + \epsilon^-_{j_2, n_2})(\epsilon^+_{j, n} +  \epsilon^-_{j_1, n_1})(\epsilon^-_{j_1, n_1} + \epsilon^+_{j^\prime, n^\prime})} \non \\
& + & \frac{(n + 1) n_2 (n_1 + 1) n^\prime } {(\epsilon^+_{j, n} + \epsilon^-_{j_2, n_2} + \epsilon^+_{j_1, n_1} +  \epsilon^-_{j^\prime, n^\prime})(\epsilon^+_{j_1, n_1} +  \epsilon^-_{j^\prime, n^\prime})(\epsilon^+_{j, n} +  \epsilon^-_{j^\prime, n^\prime})} \non \\
& +  & \frac{ n (n_2 + 1) n_1 (n^\prime + 1)} {(\epsilon^-_{j, n} + \epsilon^+_{j_2, n_2} + \epsilon^-_{j_1, n_1} +  \epsilon^+_{j^\prime, n^\prime})(\epsilon^-_{j, n} +  \epsilon^+_{j_2, n_2})(\epsilon^+_{j_2, n_2} +  \epsilon^-_{j_1, n_1})} ]  \non \\
& + & [ \frac{n n_2 (n_1 + 1) (n^\prime + 1)}{(\epsilon^-_{j, n} +  \epsilon^+_{j^\prime, n^\prime})(\epsilon^-_{j_2, n_2} +  \epsilon^+_{j_1, n_1})(\epsilon^-_{j_2, n_2} +  \epsilon^+_{j^\prime, n^\prime})} + \frac{n (n_2 + 1) (n_1 + 1) n^\prime}{(\epsilon^-_{j, n} + \epsilon^+_{j_2, n_2})(\epsilon^+_{j_2, n_2} +  \epsilon^-_{j^\prime, n^\prime})(\epsilon^+_{j_1, n_1} + \epsilon^-_{j^\prime, n^\prime})} \non \\
& + & \frac{(n + 1) n_2 (n_1 + 1) n^\prime } {(\epsilon^+_{j, n} + \epsilon^-_{j_2, n_2} + \epsilon^+_{j_1, n_1} +  \epsilon^-_{j^\prime, n^\prime})(\epsilon^+_{j, n} +  \epsilon^-_{j_2, n_2})(\epsilon^+_{j, n} +  \epsilon^-_{j^\prime, n^\prime})} \non \\
& +  & \frac{ n (n_2 + 1) n_1 (n^\prime + 1)} {(\epsilon^-_{j, n} + \epsilon^+_{j_2, n_2} + \epsilon^-_{j_1, n_1} +  \epsilon^+_{j^\prime, n^\prime})(\epsilon^+_{j_2, n_2} +  \epsilon^-_{j_1, n_1})(\epsilon^-_{j_1, n_1} + \epsilon^+_{j^\prime, n^\prime})} ] \non \\
& + & [ \frac{ n (n_2 + 1) n_1 (n^\prime + 1)} {(\epsilon^-_{j, n} + \epsilon^+_{j_2, n_2} + \epsilon^-_{j_1, n_1} +  \epsilon^+_{j^\prime, n^\prime})(\epsilon^-_{j, n} +  \epsilon^+_{j_2, n_2})(\epsilon^-_{j, n} + \epsilon^+_{j^\prime, n^\prime})}  \non \\
& + & \frac{(n + 1) n_2 (n_1 + 1) n^\prime } {(\epsilon^+_{j, n} + \epsilon^-_{j_2, n_2} + \epsilon^+_{j_1, n_1} +  \epsilon^-_{j^\prime, n^\prime})(\epsilon^-_{j_2, n_2} +  \epsilon^+_{j_1, n_1})(\epsilon^+_{j_1, n_1} +  \epsilon^-_{j^\prime, n^\prime})} \non \\
& +  & \frac{(n + 1)n_2 n_1 (n^\prime + 1)}{(\epsilon^+_{j, n} +  \epsilon^-_{j_2, n_2})(\epsilon^-_{j_2, n_2} +  \epsilon^+_{j^\prime, n^\prime})(\epsilon^-_{j_1, n_1} +  \epsilon^+_{j^\prime, n^\prime})} + \frac{(n + 1) (n_2 + 1) n_1 n^\prime} {(\epsilon^+_{j_2, n_2} + \epsilon^-_{j_1, n_1})(\epsilon^+_{j_2, n_2} +  \epsilon^-_{j^\prime, n^\prime})(\epsilon^+_{j, n} + \epsilon^-_{j^\prime, n^\prime})}] \non \\
& + & [ \frac{ n (n_2 + 1) n_1 (n^\prime + 1)} {(\epsilon^-_{j, n} + \epsilon^+_{j_2, n_2} + \epsilon^-_{j_1, n_1} +  \epsilon^+_{j^\prime, n^\prime})(\epsilon^-_{j, n} + \epsilon^+_{j^\prime, n^\prime})(\epsilon^-_{j_1, n_1} + \epsilon^+_{j^\prime, n^\prime})} \non \\
& + & \frac{(n + 1) n_2 (n_1 + 1) n^\prime } {(\epsilon^+_{j, n} + \epsilon^-_{j_2, n_2} + \epsilon^+_{j_1, n_1} +  \epsilon^-_{j^\prime, n^\prime})(\epsilon^+_{j, n} + \epsilon^-_{j_2, n_2})(\epsilon^-_{j_2, n_2} +  \epsilon^+_{j_1, n_1})} \non \\
& +  & \frac{(n + 1)(n_2 + 1) n_1 n^\prime}{(\epsilon^+_{j, n} +  \epsilon^-_{j_1, n_1})(\epsilon^+_{j_2, n_2} +  \epsilon^-_{j_1, n_1})(\epsilon^+_{j, n} +  \epsilon^-_{j^\prime, n^\prime})} + \frac{n (n_2 + 1) (n_1 + 1)n^\prime} {(\epsilon^-_{j, n} + \epsilon^+_{j_2, n_2})(\epsilon^-_{j, n} + \epsilon^+_{j_1, n_1})(\epsilon^+_{j_1, n_1} + \epsilon^-_{j^\prime, n^\prime})} ] \non \\
& + & [ \frac{ n n_2 n_1 (n^\prime + 1)} {(\epsilon^-_{j, n} + \epsilon^+_{j^\prime, n^\prime})(\epsilon^-_{j_2, n_2} + \epsilon^+_{j^\prime, n^\prime})(\epsilon^-_{j_1, n_1} + \epsilon^+_{j^\prime, n^\prime})} +  \frac{n (n_2 + 1) n_1 n^\prime} {(\epsilon^-_{j, n} + \epsilon^+_{j_2, n_2})(\epsilon^+_{j_2, n_2} + \epsilon^-_{j_1, n_1})(\epsilon^+_{j_2, n_2} + \epsilon^-_{j^\prime, n^\prime})} \non \\
& + &  \frac{(n + 1) n_2 n_1 n^\prime} {(\epsilon^+_{j, n} +  \epsilon^-_{j_2, n_2})(\epsilon^+_{j, n} +  \epsilon^-_{j_1, n_1})(\epsilon^+_{j, n} +  \epsilon^-_{j^\prime, n^\prime})} + \frac{n n_2 (n_1 + 1) n^\prime } {(\epsilon^-_{j, n} + \epsilon^+_{j_1, n_1})(\epsilon^-_{j_2, n_2} +  \epsilon^+_{j_1, n_1})(\epsilon^+_{j_1, n_1} +  \epsilon^-_{j^\prime, n^\prime})} ] \; \}, \non \\
&&
\label{eq: C3-jj2j1jp}
\eea
\newpage
\bea
&&\bar{\mathfrak{C}}^{(3)}_{j j^\prime j_1 j^\prime} = \mathfrak{t}_{j j^\prime} \mathfrak{t}_{j^\prime j_1} \mathfrak{t}_{j_1 j^\prime} \, \sum_{n, n_1, n^\prime}  \; \rho_{j, n} \,\rho_{j_1, n_1} \, \rho_{j^\prime, n^\prime}  \; \times \non \\
& \{ & [ \frac{ n (n^\prime + 1)(n_1 + 1)(n^\prime + 1)}{(\epsilon^-_{j, n} + \epsilon^+_{j^\prime, n_j^\prime})(\epsilon^-_{j, n} +  \epsilon^+_{j_1, n_1})(\epsilon^-_{j, n} + \epsilon^+_{j^\prime, n_j^\prime})}  + \frac{ (n + 1) (n^\prime + 1) n_1 (n^\prime + 1)} {(\epsilon^-_{j_1, n_1} +  \epsilon^+_{j^\prime, n^\prime })(\epsilon^+_{j, n} +  \epsilon^-_{j_1, n_1})(\epsilon^-_{j_1, n_1} +  \epsilon^+_{j^\prime, n^\prime })} \non \\
& + &  \frac{ (n + 1) n^\prime (n_1 + 1)n^\prime } {(\epsilon^+_{j, n} +  \epsilon^-_{j^\prime, n^\prime })(\epsilon^+_{j_1, n_1} + \epsilon^-_{j^\prime, n^\prime })} (\beta - \frac{1}{(\epsilon^+_{j, n} +  \epsilon^-_{j^\prime, n^\prime })} - \frac{1}{(\epsilon^+_{j_1, n_1} + \epsilon^-_{j^\prime, n^\prime })})] \non \\
& + & [ \frac{n n^\prime  (n_1 + 1) (n^\prime + 1)}{(\epsilon^-_{j, n} +  \epsilon^+_{j_1, n_1})(\epsilon^+_{j_1, n_1} +  \epsilon^-_{j^\prime, n^\prime})(\epsilon^-_{j, n} +  \epsilon^+_{j^\prime, n^\prime})} + \frac{(n + 1) n^\prime  n_1 (n^\prime + 1)}{(\epsilon^+_{j, n} + \epsilon^-_{j_1, n_1})(\epsilon^+_{j, n} +  \epsilon^-_{j^\prime, n^\prime})(\epsilon^-_{j_1, n_1} + \epsilon^+_{j^\prime, n^\prime})} \non \\
& + & \frac{(n + 1) (n^\prime - 1)(n_1 + 1) n^\prime } {(\epsilon^+_{j, n} + \epsilon^{-}_{j^\prime, n^\prime - 1} + \epsilon^+_{j_1, n_1} +  \epsilon^{-}_{j^\prime, n^\prime})(\epsilon^+_{j_1, n_1} +  \epsilon^-_{j^\prime, n^\prime})(\epsilon^+_{j, n} +  \epsilon^-_{j^\prime, n^\prime})} \non \\
& +  & \frac{ n (n^\prime + 2) n_1 (n^\prime + 1)} {(\epsilon^-_{j, n} + \epsilon^{+}_{j^\prime, n^\prime + 1} + \epsilon^-_{j_1, n_1} +  \epsilon^{+}_{j^\prime, n^\prime})(\epsilon^-_{j, n} +  \epsilon^+_{j^\prime, n^\prime})(\epsilon^-_{j_1, n_1} +  \epsilon^+_{j^\prime, n^\prime})} ]  \non \\
& + & [ \frac{n n^\prime (n_1 + 1) (n^\prime + 1)}{(\epsilon^-_{j, n} +  \epsilon^+_{j^\prime, n^\prime})(\epsilon^+_{j_1, n_1} + \epsilon^-_{j^\prime, n^\prime})} ( \beta - \frac{1}{(\epsilon^-_{j, n} +  \epsilon^+_{j^\prime, n^\prime})} - \frac{1}{(\epsilon^+_{j_1, n_1} + \epsilon^-_{j^\prime, n^\prime})} )  \non \\
& + & \frac{(n + 1) (n^\prime - 1) (n_1 + 1)  n^\prime } {(\epsilon^+_{j, n}  + \epsilon^{-}_{j^\prime, n^\prime - 1} + \epsilon^+_{j_1, n_1} +  \epsilon^{-}_{j^\prime, n^\prime})(\epsilon^+_{j, n} +  \epsilon^-_{j^\prime, n^\prime})(\epsilon^+_{j, n} +  \epsilon^-_{j^\prime, n^\prime})} \non \\
& +  & \frac{ n (n^\prime + 1)  n_1 (n^\prime + 2)} {(\epsilon^-_{j, n} + \epsilon^+_{j^\prime, n^\prime } + \epsilon^-_{j_1, n_1} +  \epsilon^+_{j^\prime, n^\prime + 1})(\epsilon^+_{j^\prime, n^\prime} +  \epsilon^-_{j_1, n_1})(\epsilon^-_{j_1, n_1} + \epsilon^+_{j^\prime, n^\prime})} ] \non \\
& + & [ \frac{ n (n^\prime + 2) n_1 (n^\prime + 1)} {(\epsilon^-_{j, n} + \epsilon^+_{j^\prime, n^\prime + 1} + \epsilon^-_{j_1, n_1} +  \epsilon^+_{j^\prime, n^\prime})(\epsilon^-_{j, n} +  \epsilon^+_{j^\prime, n^\prime})(\epsilon^-_{j, n} + \epsilon^+_{j^\prime, n^\prime})}  \non \\
& + & \frac{(n + 1) (n^\prime-1) (n_1 + 1) n^\prime } {(\epsilon^+_{j, n} + \epsilon^-_{j^\prime, n^\prime -1} + \epsilon^+_{j_1, n_1} +  \epsilon^-_{j^\prime, n^\prime})(\epsilon^-_{j^\prime, n^\prime} +  \epsilon^+_{j_1, n_1})(\epsilon^+_{j_1, n_1} +  \epsilon^-_{j^\prime, n^\prime})} \non \\
& +  & \frac{(n + 1) n^\prime n_1 (n^\prime + 1)}{(\epsilon^+_{j, n} +  \epsilon^-_{j^\prime, n^\prime})(\epsilon^-_{j_1, n_1} +  \epsilon^+_{j^\prime, n^\prime})}(\beta -\frac{1}{(\epsilon^+_{j, n} +  \epsilon^-_{j^\prime, n^\prime})} -\frac{1}{(\epsilon^-_{j_1, n_1} +  \epsilon^+_{j^\prime, n^\prime})}) ] \non \\
& + & [ \frac{ n (n^\prime + 2) n_1 (n^\prime + 1)} {(\epsilon^-_{j, n} + \epsilon^+_{j^\prime, n^\prime + 1} + \epsilon^-_{j_1, n_1} +  \epsilon^+_{j^\prime, n^\prime})(\epsilon^-_{j, n} + \epsilon^+_{j^\prime, n^\prime})(\epsilon^-_{j_1, n_1} + \epsilon^+_{j^\prime, n^\prime})} \non \\
& + & \frac{(n + 1) n^\prime (n_1 + 1) (n^\prime - 1) } {(\epsilon^+_{j, n} + \epsilon^-_{j^\prime, n^\prime} + \epsilon^+_{j_1, n_1} +  \epsilon^-_{j^\prime, n^\prime - 1})(\epsilon^+_{j, n} + \epsilon^-_{j^\prime, n^\prime})(\epsilon^-_{j^\prime, n^\prime} +  \epsilon^+_{j_1, n_1})} \non \\
& +  & \frac{(n + 1)(n^\prime + 1) n_1 n^\prime}{(\epsilon^+_{j, n} +  \epsilon^-_{j_1, n_1})(\epsilon^+_{j^\prime, n^\prime} +  \epsilon^-_{j_1, n_1})(\epsilon^+_{j, n} +  \epsilon^-_{j^\prime, n^\prime})} + \frac{n (n^\prime + 1) (n_1 + 1)n^\prime} {(\epsilon^-_{j, n} + \epsilon^+_{j^\prime, n^\prime})(\epsilon^-_{j, n} + \epsilon^+_{j_1, n_1})(\epsilon^+_{j_1, n_1} + \epsilon^-_{j^\prime, n^\prime})} ] \non \\
& + & [ \frac{ n (n^\prime + 1) n_1 (n^\prime + 1)} {(\epsilon^-_{j, n} + \epsilon^+_{j^\prime, n^\prime})(\epsilon^-_{j_1, n_1} + \epsilon^+_{j^\prime, n^\prime})}(\beta - \frac{1}{(\epsilon^-_{j, n} + \epsilon^+_{j^\prime, n^\prime})} - \frac{1}{(\epsilon^-_{j_1, n_1} + \epsilon^+_{j^\prime, n^\prime})}) \non \\
& + & \frac{(n + 1) n^\prime n_1 n^\prime} {(\epsilon^+_{j, n} +  \epsilon^-_{j^\prime, n^\prime})(\epsilon^+_{j, n} +  \epsilon^-_{j_1, n_1})(\epsilon^+_{j, n} +  \epsilon^-_{j^\prime, n^\prime})} + \frac{n n^\prime (n_1 + 1) n^\prime } {(\epsilon^-_{j, n} + \epsilon^+_{j_1, n_1})(\epsilon^-_{j^\prime, n^\prime} +  \epsilon^+_{j_1, n_1})(\epsilon^+_{j_1, n_1} +  \epsilon^-_{j^\prime, n^\prime})} ] \; \}. \non \\
&&
\label{eq: C3-jjpj1jp}
\eea
Similarly, we get
\bea
&&\bar{\mathfrak{C}}^{(3)}_{j j_2 j j^\prime}  =  \mathfrak{t}_{j j_2} \mathfrak{t}_{j_2 j} \mathfrak{t}_{j j^\prime} \, \sum_{n, n_2, n^\prime}  \; \rho_{j, n} \, \rho_{j_2, n_2} \, \rho_{j^\prime, n^\prime}  \; \times \non \\
& \{ & [ \frac{ n (n_2 + 1) n (n^\prime + 1)}{(\epsilon^-_{j, n} +  \epsilon^+_{j_2, n_2})(\epsilon^-_{j, n} + \epsilon^+_{j^\prime, n_j^\prime})} (\beta - \frac{1}{(\epsilon^-_{j, n} +  \epsilon^+_{j_2, n_2})} - \frac{1}{(\epsilon^-_{j, n} + \epsilon^+_{j^\prime, n_j^\prime})})  \non \\
& + &  \frac{ (n + 1) (n_2 + 1)(n + 1) n^\prime } {(\epsilon^+_{j, n} +  \epsilon^-_{j^\prime, n^\prime}) (\epsilon^+_{j_2, n_2} +  \epsilon^-_{j^\prime, n^\prime})(\epsilon^+_{j, n} +  \epsilon^-_{j^\prime, n^\prime})}  + \frac{ (n + 1) n_2 (n + 1)  (n^\prime + 1)} {(\epsilon^+_{j, n} +  \epsilon^-_{j_2, n_2})(\epsilon^+_{j, n} +  \epsilon^-_{j_2, n_2})(\epsilon^-_{j_2, n_2} +  \epsilon^+_{j^\prime, n^\prime })} ] \non \\
& + & [ \frac{n n_2 (n + 1) (n^\prime + 1)}{(\epsilon^-_{j, n} +  \epsilon^+_{j^\prime, n^\prime})(\epsilon^-_{j_2, n_2} +  \epsilon^+_{j, n})} (\beta - \frac{1}{(\epsilon^-_{j, n} +  \epsilon^+_{j^\prime, n^\prime})} -  \frac{1}{(\epsilon^-_{j_2, n_2} +  \epsilon^+_{j, n})}) \non \\
& + & \frac{(n + 1) n_2 (n + 2) n^\prime } {(\epsilon^+_{j, n} + \epsilon^-_{j_2, n_2} + \epsilon^+_{j, n + 1} +  \epsilon^-_{j^\prime, n^\prime})(\epsilon^+_{j, n} +  \epsilon^-_{j^\prime, n^\prime})(\epsilon^+_{j, n} +  \epsilon^-_{j^\prime, n^\prime})} \non \\
& +  & \frac{ n (n_2 + 1) (n - 1) (n^\prime + 1)} {(\epsilon^-_{j, n} + \epsilon^+_{j_2, n_2} + \epsilon^-_{j, n - 1} +  \epsilon^+_{j^\prime, n^\prime})(\epsilon^-_{j, n} +  \epsilon^+_{j_2, n_2})(\epsilon^+_{j_2, n_2} +  \epsilon^-_{j, n})} ]  \non \\
& + & [ \frac{n n_2 (n + 1) (n^\prime + 1)}{(\epsilon^-_{j, n} +  \epsilon^+_{j^\prime, n^\prime})(\epsilon^-_{j_2, n_2} +  \epsilon^+_{j, n})(\epsilon^-_{j_2, n_2} +  \epsilon^+_{j^\prime, n^\prime})} + \frac{n (n_2 + 1) (n + 1) n^\prime}{(\epsilon^-_{j, n} + \epsilon^+_{j_2, n_2})(\epsilon^+_{j_2, n_2} +  \epsilon^-_{j^\prime, n^\prime})(\epsilon^+_{j, n} + \epsilon^-_{j^\prime, n^\prime})} \non \\
& + & \frac{(n + 1) n_2 (n + 2) n^\prime } {(\epsilon^+_{j, n} + \epsilon^-_{j_2, n_2} + \epsilon^+_{j, n + 1} +  \epsilon^-_{j^\prime, n^\prime})(\epsilon^+_{j, n} +  \epsilon^-_{j_2, n_2})(\epsilon^+_{j, n} +  \epsilon^-_{j^\prime, n^\prime})} \non \\
& +  & \frac{ n (n_2 + 1) (n - 1) (n^\prime + 1)} {(\epsilon^-_{j, n} + \epsilon^+_{j_2, n_2} + \epsilon^-_{j, n - 1} +  \epsilon^+_{j^\prime, n^\prime})(\epsilon^+_{j_2, n_2} +  \epsilon^-_{j, n})(\epsilon^-_{j, n} + \epsilon^+_{j^\prime, n^\prime})} ] \non \\
& + & [ \frac{ n (n_2 + 1) (n - 1) (n^\prime + 1)} {(\epsilon^-_{j, n} + \epsilon^+_{j_2, n_2} + \epsilon^-_{j, n - 1} +  \epsilon^+_{j^\prime, n^\prime})(\epsilon^-_{j, n} +  \epsilon^+_{j_2, n_2})(\epsilon^-_{j, n} + \epsilon^+_{j^\prime, n^\prime})}  \non \\
& + & \frac{(n + 1) n_2 (n + 2) n^\prime } {(\epsilon^+_{j, n} + \epsilon^-_{j_2, n_2} + \epsilon^+_{j, n + 1} +  \epsilon^-_{j^\prime, n^\prime})(\epsilon^-_{j_2, n_2} +  \epsilon^+_{j, n})(\epsilon^+_{j, n} +  \epsilon^-_{j^\prime, n^\prime})} \non \\
& +  &  \frac{(n + 1) (n_2 + 1) n n^\prime} {(\epsilon^+_{j_2, n_2} + \epsilon^-_{j, n})(\epsilon^+_{j_2, n_2} +  \epsilon^-_{j^\prime, n^\prime})(\epsilon^+_{j, n} + \epsilon^-_{j^\prime, n^\prime})} + \frac{(n + 1) n_2 n (n^\prime + 1)} {(\epsilon^+_{j, n} +  \epsilon^-_{j_2, n_2})(\epsilon^-_{j_2, n_2} +  \epsilon^+_{j^\prime, n^\prime})(\epsilon^-_{j, n} +  \epsilon^+_{j^\prime, n^\prime})} ] \non \\
& + & [ \frac{ n (n_2 + 1) (n-1) (n^\prime + 1)} {(\epsilon^-_{j, n} + \epsilon^+_{j_2, n_2} + \epsilon^-_{j, n-1} +  \epsilon^+_{j^\prime, n^\prime})(\epsilon^-_{j, n} + \epsilon^+_{j^\prime, n^\prime})(\epsilon^-_{j, n} + \epsilon^+_{j^\prime, n^\prime})} \non \\
& + & \frac{(n + 1) n_2 (n + 2) n^\prime } {(\epsilon^+_{j, n} + \epsilon^-_{j_2, n_2} + \epsilon^+_{j, n + 1} +  \epsilon^-_{j^\prime, n^\prime})(\epsilon^+_{j, n} + \epsilon^-_{j_2, n_2})(\epsilon^-_{j_2, n_2} +  \epsilon^+_{j, n})} \non \\
& +  & \frac{(n + 1)(n_2 + 1) n n^\prime}{(\epsilon^+_{j_2, n_2} +  \epsilon^-_{j, n})(\epsilon^+_{j, n} +  \epsilon^-_{j^\prime, n^\prime})}(\beta - \frac{1}{(\epsilon^+_{j_2, n_2} +  \epsilon^-_{j, n})} - \frac{1}{(\epsilon^+_{j, n} +  \epsilon^-_{j^\prime, n^\prime})}) ] \non \\
& + & [ \frac{ n n_2 n (n^\prime + 1)} {(\epsilon^-_{j, n} + \epsilon^+_{j^\prime, n^\prime})(\epsilon^-_{j_2, n_2} + \epsilon^+_{j^\prime, n^\prime})(\epsilon^-_{j, n} + \epsilon^+_{j^\prime, n^\prime})} + \frac{n (n_2 + 1) n n^\prime} {(\epsilon^-_{j, n} + \epsilon^+_{j_2, n_2})(\epsilon^+_{j_2, n_2} + \epsilon^-_{j, n})(\epsilon^+_{j_2, n_2} + \epsilon^-_{j^\prime, n^\prime})}  \non \\
& + & \frac{(n + 1) n_2 (n + 1) n^\prime} {(\epsilon^+_{j, n} +  \epsilon^-_{j_2, n_2})(\epsilon^+_{j, n} +  \epsilon^-_{j^\prime, n^\prime})} (\beta - \frac{1}{(\epsilon^+_{j, n} +  \epsilon^-_{j_2, n_2})} - \frac{1}{(\epsilon^+_{j, n} +  \epsilon^-_{j^\prime, n^\prime})}) ] \; \} \non \\
&&
\label{eq: C3-jj2jjp}
\eea
and
\bea
&& \bar{\mathfrak{C}}^{(3)}_{j j^\prime j j^\prime} = \mathfrak{t}_{j j^\prime} \mathfrak{t}_{j^\prime j} \mathfrak{t}_{j j^\prime} \, \sum_{n, n^\prime}  \; \rho_{j, n} \, \rho_{j^\prime, n^\prime}  \; \times \non \\
& \{ & [ \frac{ n (n^\prime + 1) n (n^\prime + 1)} {(\epsilon^-_{j, n} + \epsilon^+_{j^\prime, n_j^\prime})(\epsilon^-_{j, n} + \epsilon^+_{j^\prime, n_j^\prime})} (\beta - \frac{2}{(\epsilon^-_{j, n} + \epsilon^+_{j^\prime, n_j^\prime})} )  \non \\
& + &  \frac{ (n + 1) n^\prime (n + 1) n^\prime } {(\epsilon^+_{j, n} +  \epsilon^-_{j^\prime, n^\prime })(\epsilon^+_{j, n} + \epsilon^-_{j^\prime, n^\prime })} (\beta - \frac{2}{(\epsilon^+_{j, n} +  \epsilon^-_{j^\prime, n^\prime })} ) ] \non \\
& + & [ \frac{n n^\prime  (n + 1) (n^\prime + 1)}{(\epsilon^+_{j, n} +  \epsilon^-_{j^\prime, n^\prime})(\epsilon^-_{j, n} +  \epsilon^+_{j^\prime, n^\prime})}(\beta - \frac{1}{(\epsilon^-_{j, n} +  \epsilon^+_{j^\prime, n^\prime})} - \frac{1}{(\epsilon^+_{j, n} +  \epsilon^-_{j^\prime, n^\prime})}) \non \\
& + & \frac{(n + 1) (n^\prime - 1)(n + 2) n^\prime } {(\epsilon^+_{j, n} + \epsilon^{-}_{j^\prime, n^\prime - 1} + \epsilon^+_{j, n + 1} +  \epsilon^{-}_{j^\prime, n^\prime})(\epsilon^+_{j, n} +  \epsilon^-_{j^\prime, n^\prime - 1})(\epsilon^+_{j, n} +  \epsilon^-_{j^\prime, n^\prime})} \non \\
& +  & \frac{ n (n^\prime + 2) (n - 1) (n^\prime + 1)} {(\epsilon^-_{j, n} + \epsilon^{+}_{j^\prime, n^\prime + 1} + \epsilon^-_{j, n - 1} +  \epsilon^{+}_{j^\prime, n^\prime})(\epsilon^-_{j, n} +  \epsilon^+_{j^\prime, n^\prime})(\epsilon^-_{j, n} +  \epsilon^+_{j^\prime, n^\prime})} ]  \non \\
& + & [ \frac{n n^\prime (n + 1) (n^\prime + 1)}{(\epsilon^-_{j, n} +  \epsilon^+_{j^\prime, n^\prime})(\epsilon^+_{j, n} + \epsilon^-_{j^\prime, n^\prime})} ( \beta - \frac{1}{(\epsilon^-_{j, n} +  \epsilon^+_{j^\prime, n^\prime})} - \frac{1}{(\epsilon^+_{j, n} + \epsilon^-_{j^\prime, n^\prime})} )  \non \\
& + & \frac{(n + 1) (n^\prime - 1) (n + 2)  n^\prime } {(\epsilon^+_{j, n}  + \epsilon^{-}_{j^\prime, n^\prime - 1} + \epsilon^+_{j, n + 1} +  \epsilon^{-}_{j^\prime, n^\prime})(\epsilon^+_{j, n} +  \epsilon^-_{j^\prime, n^\prime})(\epsilon^+_{j, n} +  \epsilon^-_{j^\prime, n^\prime})} \non \\
& +  & \frac{ n (n^\prime + 1)  (n - 1) (n^\prime + 2)} {(\epsilon^-_{j, n} + \epsilon^+_{j^\prime, n^\prime } + \epsilon^-_{j, n - 1} +  \epsilon^+_{j^\prime, n^\prime + 1})(\epsilon^+_{j^\prime, n^\prime} +  \epsilon^-_{j, n})(\epsilon^-_{j, n} + \epsilon^+_{j^\prime, n^\prime})} ] \non \\
& + & [ \frac{ n (n^\prime + 2) (n - 1) (n^\prime + 1)} {(\epsilon^-_{j, n} + \epsilon^+_{j^\prime, n^\prime + 1} + \epsilon^-_{j, n - 1} +  \epsilon^+_{j^\prime, n^\prime})(\epsilon^-_{j, n} +  \epsilon^+_{j^\prime, n^\prime})(\epsilon^-_{j, n} + \epsilon^+_{j^\prime, n^\prime})}  \non \\
& + & \frac{(n + 1) (n^\prime-1) (n + 2) n^\prime } {(\epsilon^+_{j, n} + \epsilon^-_{j^\prime, n^\prime -1} + \epsilon^+_{j, n + 1} +  \epsilon^-_{j^\prime, n^\prime})(\epsilon^-_{j^\prime, n^\prime} +  \epsilon^+_{j, n})(\epsilon^+_{j, n} +  \epsilon^-_{j^\prime, n^\prime})} \non \\
& +  & \frac{(n + 1) n^\prime n (n^\prime + 1)}{(\epsilon^+_{j, n} +  \epsilon^-_{j^\prime, n^\prime})(\epsilon^-_{j, n} +  \epsilon^+_{j^\prime, n^\prime})}(\beta -\frac{1}{(\epsilon^+_{j, n} +  \epsilon^-_{j^\prime, n^\prime})} -\frac{1}{(\epsilon^-_{j, n} +  \epsilon^+_{j^\prime, n^\prime})}) ] \non \\
& + & [ \frac{ n (n^\prime + 2) (n - 1) (n^\prime + 1)} {(\epsilon^-_{j, n} + \epsilon^+_{j^\prime, n^\prime + 1} + \epsilon^-_{j, n - 1} +  \epsilon^+_{j^\prime, n^\prime})(\epsilon^-_{j, n} + \epsilon^+_{j^\prime, n^\prime})(\epsilon^-_{j, n} + \epsilon^+_{j^\prime, n^\prime})} \non \\
& + & \frac{(n + 1) (n^\prime - 1) (n + 2) n^\prime} {(\epsilon^+_{j, n} + \epsilon^-_{j^\prime, n^\prime - 1} + \epsilon^+_{j, n + 1} +  \epsilon^-_{j^\prime, n^\prime})(\epsilon^+_{j, n} + \epsilon^-_{j^\prime, n^\prime})(\epsilon^-_{j^\prime, n^\prime} +  \epsilon^+_{j_1, n_1})} \non \\
& +  & \frac{(n + 1)(n^\prime + 1) n n^\prime}{(\epsilon^+_{j^\prime, n^\prime} +  \epsilon^-_{j, n})(\epsilon^+_{j, n} +  \epsilon^-_{j^\prime, n^\prime})} (\beta - \frac{1}{(\epsilon^+_{j^\prime, n^\prime} +  \epsilon^-_{j, n})} - \frac{1}{(\epsilon^+_{j, n} +  \epsilon^-_{j^\prime, n^\prime})})] \non \\
& + & [ \frac{ n (n^\prime + 1) n (n^\prime + 1)} {(\epsilon^-_{j, n} + \epsilon^+_{j^\prime, n^\prime})(\epsilon^-_{j, n} + \epsilon^+_{j^\prime, n^\prime})}(\beta - \frac{2}{(\epsilon^-_{j, n} + \epsilon^+_{j^\prime, n^\prime})} ) \non \\
& + & \frac{(n + 1) n^\prime (n + 1) n^\prime} {(\epsilon^+_{j, n} +  \epsilon^-_{j^\prime, n^\prime})(\epsilon^+_{j, n} +  \epsilon^-_{j^\prime, n^\prime})}(\beta - \frac{2}{(\epsilon^+_{j, n} +  \epsilon^-_{j^\prime, n^\prime})})
] \; \}. \non \\
&&
\label{eq: C3-jjpjjp}
\eea
The symmetry of the various terms in, and the term by term correspondence between, Eqs.~(\ref{eq: C3-jj2j1jp}--\ref{eq: C3-jjpjjp}) above are noteworthy.
The above results are sufficient for the purposes of this paper, where we discuss only bipartite lattices (specifically, hypercubic lattices in $d$ dimensions with nearest-neighbor hopping only).


\begin{thebibliography}{99}
\bibitem{sf-expn}
In order to generate expansions about regions with superfluidity, one needs to include a break up of the hopping term that allows for nonzero values of the expectation values $\langle a^{\dagger}_{j} \rangle$  and $\langle a_j \rangle$, as in mean-field treatment of the Bose Hubbard model~\cite{sheshadri_etal}, into the relevant piece of $\mathcal{H}_{0 j}$.
The resulting terms are more complicated than the ones discussed here, involving anomalous multiparticle single-site Green's functions (including those with an odd number of creation/destruction operators). We hope to discuss these types of terms in future work.
\end{thebibliography}
\end{document}